\documentclass[]{pasj00}

\begin{document}
\SetRunningHead{Author(s) in page-head}{Running Head}
\Received{2009/12/17}
\Accepted{2010/11/02}

\title{First Detection of Ar-K Line Emission \\ from the Cygnus Loop}

\author{Hiroyuki \textsc{Uchida}\altaffilmark{1}, Hiroshi \textsc{Tsunemi}\altaffilmark{1}, Nozomu \textsc{Tominaga}\altaffilmark{2,3}, Satoru \textsc{Katsuda}\altaffilmark{4}, Masashi \textsc{Kimura}\altaffilmark{1}, Hiroko \textsc{Kosugi}\altaffilmark{1}, Hiroaki \textsc{Takahashi}\altaffilmark{1}, and Satoru \textsc{Takakura}\altaffilmark{1}} %
\altaffiltext{1}{Department of Earth and Space Science, Graduate School of
  Science, Osaka University, Toyonaka, Osaka 560-0043, Japan}
\altaffiltext{2}{Department of Physics, Faculty of Science and Engineering,
Konan University, Kobe, Hyogo 658-8501, Japan}
\altaffiltext{3}{Institute for the Physics and Mathematics of the Universe, 
 University of Tokyo, 5-1-5 Kashiwanoha, Kashiwa, Chiba, 277-8569, Japan}
\altaffiltext{4}{Code 662, NASA Goddard Space Flight Center, Greenbelt, MD 20771, USA}
\email{uchida@ess.sci.osaka-u.ac.jp}


\maketitle

\begin{abstract}
We observed the Cygnus Loop with \textit{XMM-Newton} (9 pointings) and \textit{Suzaku} (32 pointings) between 2002 and 2008.
The total effective exposure time is 670.2 ks.
By using all the available data, we intended to improve a signal-to-noise ratio of the spectrum.
Accordingly, the accumulated spectra obtained by the XIS and the EPIC show some line features around 3 keV which are attributed to S He$\beta$ and Ar He$\alpha$ lines, respectively.
Since the Cygnus Loop is an evolved ($\sim$10,000 yr) supernova remnant whose temperature is relatively low ($<$1 keV) compared with other young remnants, its spectrum is generally faint above 3.0 keV, no emission lines such as Ar-K line have ever been detected.
The detection of Ar-K line is the first time and we found that its abundance is significantly higher than that of the solar value; 9.0$^{+4.0}_{-3.8}$ and 8.4$^{+2.5}_{-2.7}$ (in units of solar) estimated from the XIS and the EPIC spectra, respectively.
We conclude that the Ar-K line is originated from the ejecta of the Cygnus Loop.
Follow-up X-ray observations to tightly constrain the abundances of Ar-rich ejecta will be useful to accurately estimate the progenitor's mass.
\end{abstract}

\section{Introduction}
The X-ray spectra from the supernova remnants (SNRs) show various emission lines of heavy elements blown off by the supernova (SN) explosions.
Since these elements are generated by the nucleosynthesis process inside the progenitor stars, the metal abundance pattern provides us with a clue to obtaining the information about the type of the SN explosion and the condition of the progenitor star.

The Cygnus Loop is one of the largest ($\sim1^\circ.4$ in radius; \cite{Levenson97}) and the brightest SNRs in the X-ray sky.
It is close to us (540$^{+100}_{-80}$ pc; \cite{Blair05}) and the age is estimated to be $\sim$10,000 yr.
Although the SN type of the Cygnus Loop is still unclear, \citet{McCray79} proposed that the SN explosion had occurred in a preexisting cavity and some other studies also support this result (e.g., \cite{Hester86}; \cite{Hester94}; \cite{Levenson97}).
\citet{Hester94} observed the Balmer-dominated northeast rim of the Loop and showed the blast wave was decelerated from $\sim$400 km s$^{-1}$ to less than 200 km s$^{-1}$ in the last 1,000 yr, which suggests that the blast wave is now propagating into the cavity wall.
Since the Cygnus Loop is an evolved shell-like SNR, the spectra from the rim regions should mainly consist of the shock-heated cavity material.
Then, by using optical and X-ray data obtained from the whole rim region of the Loop, \citet{Levenson98} calculated the size of the cavity wall and estimated the progenitor star to be B0, a $\sim$15\MO \ star.
This result supports the origin of the Cygnus Loop to be a core-collapse explosion rather than the Type Ia explosion.

Meanwhile, based on the analysis of the ejecta emission in X-rays, some previous studies also suggested that the Cygnus Loop is originated from a massive star of 12-15\MO \ \citep{Tsunemi07, Kimura09, Uchida09a}.
\citet{Tsunemi07} observed this SNR along the diameter from the northeast to the southwest with the \textit{XMM-Newton} satellite and showed the X-ray spectra consist of two components with different temperatures. 
They interpreted this result as follows: the low-temperature component originating from the cavity wall surrounds the high-temperature ejecta component. 
Comparing the metal abundances (Ne, Mg, Si, S, and Fe) of the ejecta component with the theoretical SN models, they concluded that the Cygnus Loop is originated from the 15\MO \ star. 
\citet{Kimura09} and \citet{Uchida09a} observed other regions of the Loop and also inferred the progenitor mass to be 12-15\MO.
These results are consistent with the estimation by \citet{Levenson98}. 
Therefore the origin of the Cygnus Loop is most likely to be a core-collapse explosion of a 12-15\MO \ star.
However, \citet{Tsunemi07}, \citet{Kimura09}, and \citet{Uchida09a} show that the number density of Fe (relative to that of O) is $\sim$10 times higher than that of the core-collapse model.
The other problem is that the central compact object such as a neutron star has not yet been detected.
If the origin of the Cygnus Loop is a core-collapse explosion, these points still remain open questions.

In this paper, we present the result of the spectral analysis of the Cygnus Loop by using all the available data obtained by the \textit{Suzaku} and the \textit{XMM-Newton} satellites.
The X-ray spectra show many emission lines, including the first detection of an Ar-K emission line from the Cygnus Loop.

\begin{figure}[t]
  \begin{center}
    \FigureFile(100mm,100mm){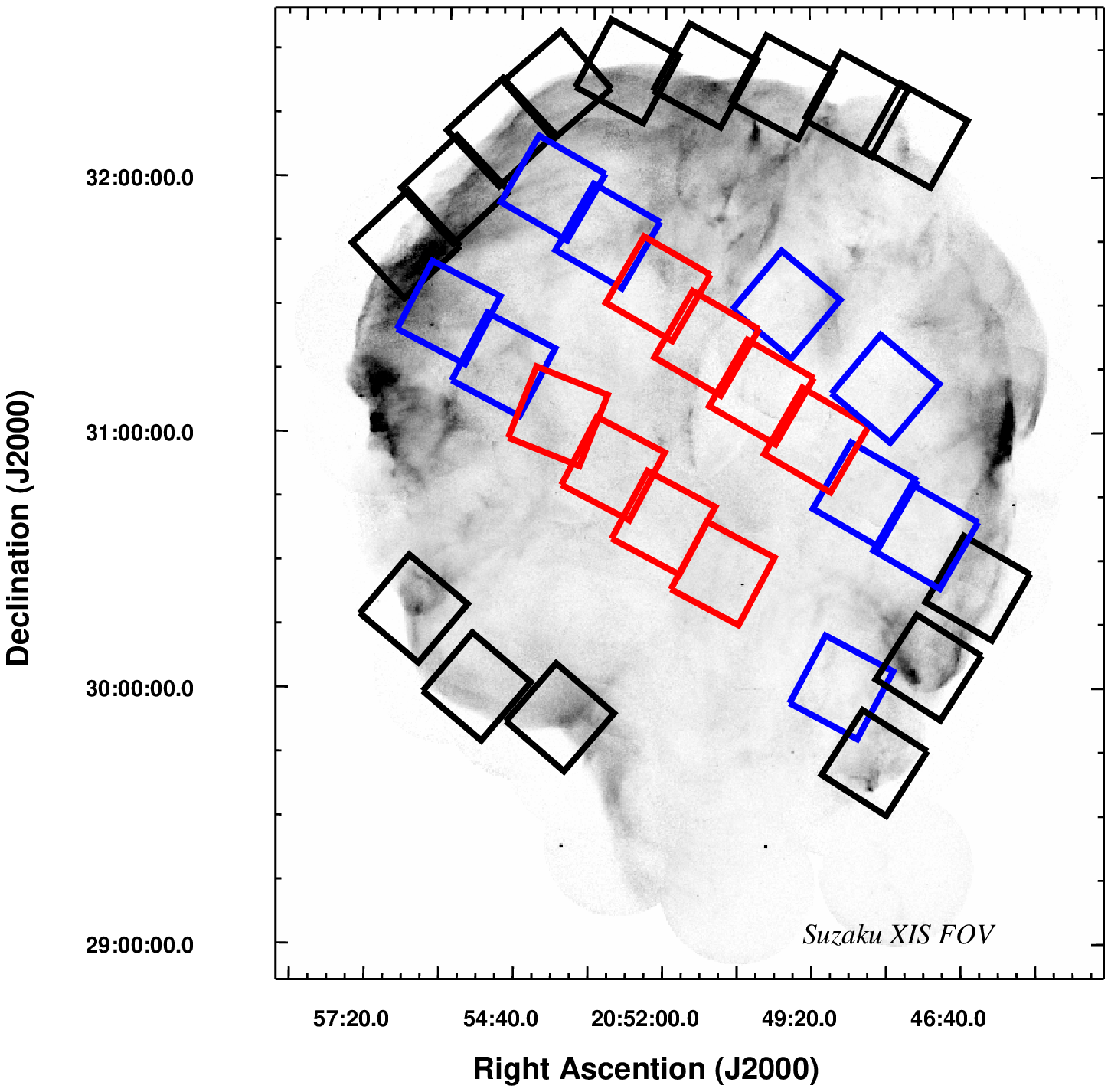}
    \FigureFile(100mm,100mm){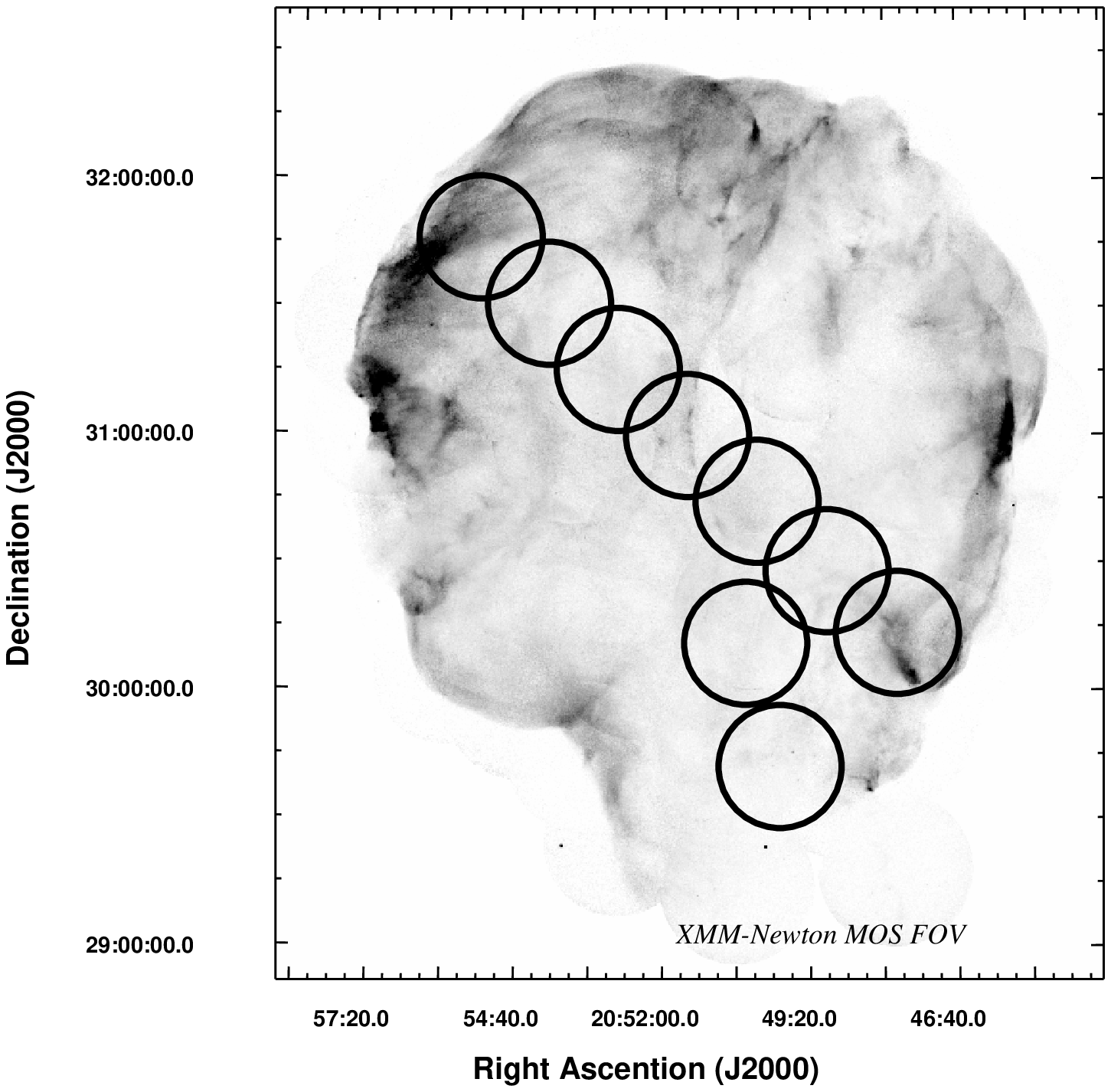}
  \end{center}
  \caption{\textit{ROSAT} HRI images of the Cygnus Loop (0.1-2.0 keV). The FOV of \textit{Suzaku} XIS and \textit{XMM-Newton} MOS are shown with solid rectangles (top) and circles (bottom), respectively. In top panel, red, blue, and black correspond to Region-A, Region-B, and Region-C, respectively.}\label{fig:HRI}
\end{figure}

\section{Observations}
The Cygnus Loop has been observed many times using X-ray CCDs.
We summarized the 41 observations in table \ref{tab:sum_suzaku} (for \textit{Suzaku}) and table \ref{tab:sum_xmm} (for \textit{XMM-Newton}). 
All the data were taken between 2002 and 2008. 
Their fields of view (FOV) are shown in figure \ref{fig:HRI}. Top and bottom panels show the FOV of the \textit{Suzaku} XIS \citep{Koyama07} and the \textit{XMM-Newton} MOS \citep{Turner01}, respectively. 

All of the \textit{Suzaku} data were analyzed with version 6.6.2 of the HEAsoft tools. For the reduction of the \textit{Suzaku} data, we used the version 9 of the Suzaku Software. The calibration database (CALDB) used was the one updated in May 2009. We used the revision 2.2 of the cleaned event data and combined the 3$\times$3 and 5$\times$5 event files. The data obtained after November 2007 (Obs. ID: 501017010) were all taken by using the spaced row charge injection (SCI) method \citep{Prigozhin08} that reduces the effect of radiation damage of the XIS and recovers the energy resolution. 
In order to exclude the background flare events, we obtained the good time intervals (GTIs) by including only times at which the count rates are within $\pm3\sigma$ of the mean count rates.
We examined both a local background  and a blank-sky data for the background subtraction.
The local background spectrum is extracted from the entire source-free regions on the periphery of the Loop  while for the blank-sky spectrum, we accumulated all of the Lockman Hole data obtained since the launch of \textit{Suzaku}.
We confirmed that there is no significant difference between two results. 
However, the total count in the Lockman Hole is an order of magnitude larger than that in the local source-free region.
We therefore prefer the Lockman Hole background to the local background.
We note that the position of the Cygnus Loop ($l = 74^\circ$, $b = -8^\circ.6$) appears to be beyond the Galactic Ridge X-ray Emission (GRXE; \cite{Sugizaki01}; \cite{Revnivtsev06}).
The solar wind charge exchange (SWCX) is also considered to be one of the causes of the soft X-ray background below 1 keV  \citep{Fujimoto07}. 
However, in terms of the Cygnus Loop, the SWCX is also negligible because of the prominent surface brightness in the low-energy band. 
Thus, for the background subtraction of the \textit{Suzaku} data, we applied the Lockman Hole data.
These observation dates were close to those of the Cygnus Loop observations and we confirmed that they have no background flares. 

All of the \textit{XMM-Newton} data were processed with version 7.1.0 of the \textit{XMM} Science Analysis System (SAS). The current calibration files (CCFs) used were the one updated on June 2008. We used the data obtained with the EPIC MOS and pn cameras. These data were taken by using the medium filters and the prime full-window mode. We selected X-ray events corresponding to patterns 0-12 and flag = 0 for MOS 1 and 2, patterns 0-4 and flag = 0 for pn, respectively.
In order to exclude the background flare events of the \textit{XMM-Newton} data, we determined the GTIs in the same way as the \textit{Suzaku} data. After filtering the data, they were vignetting-corrected using the SAS task \textit{evigweight}. For the background subtraction, we employed a blank-sky observations prepared by \citet{Read03} for the similar reason with the \textit{Suzaku}'s case. 

\begin{figure}[t]
  \begin{center}
    \FigureFile(70mm,70mm){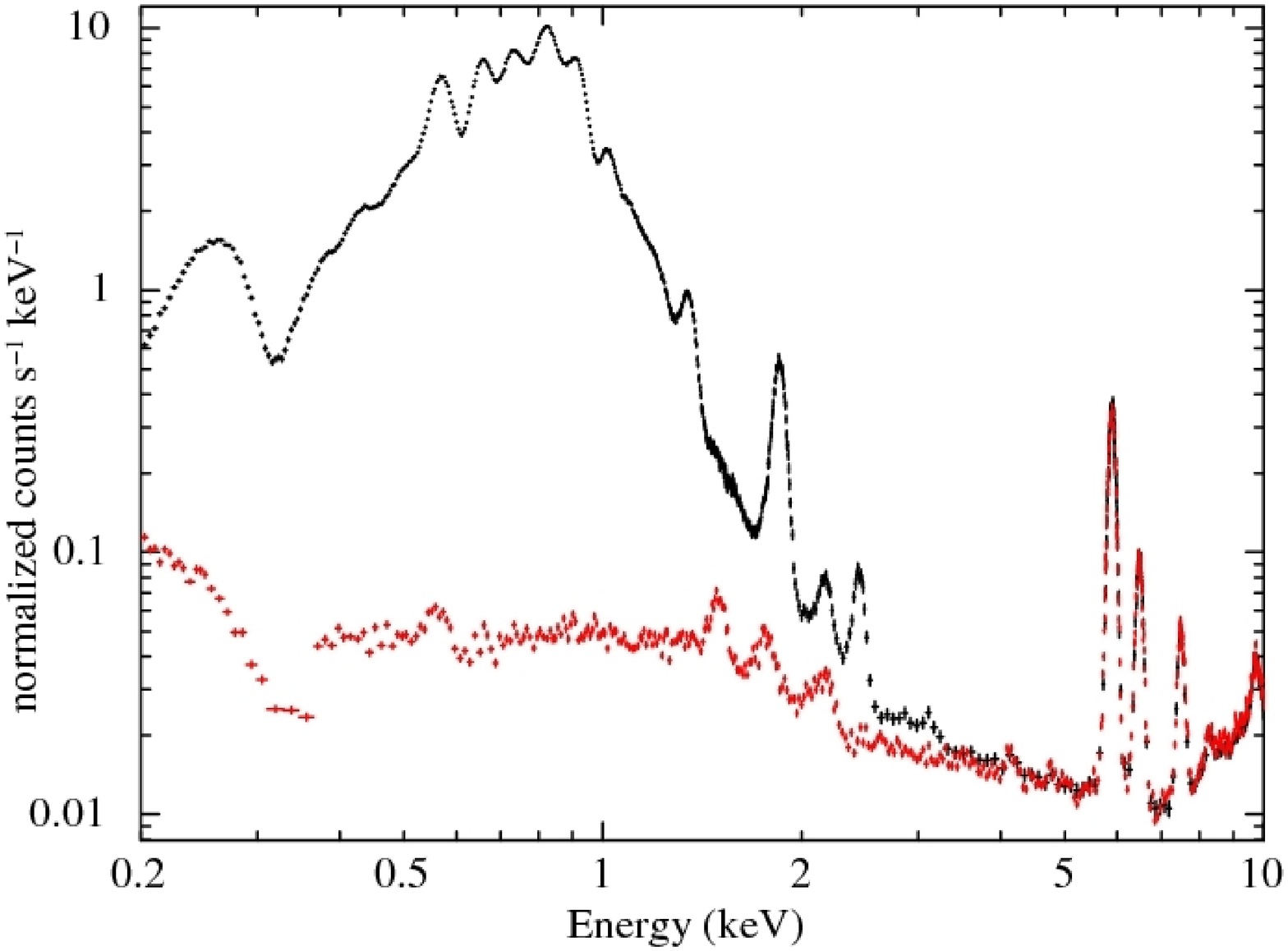}
    \FigureFile(70mm,70mm){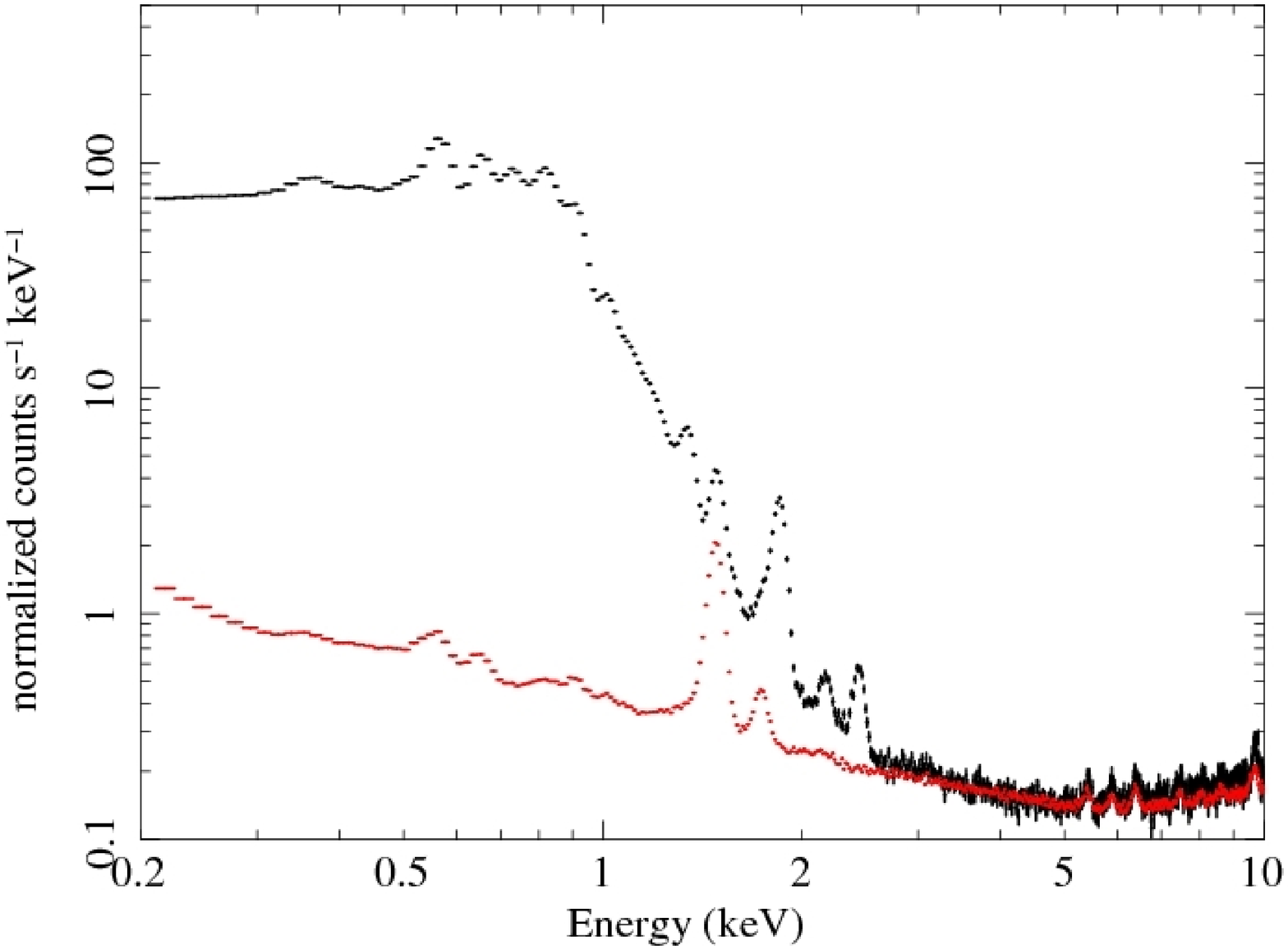}
    \FigureFile(70mm,70mm){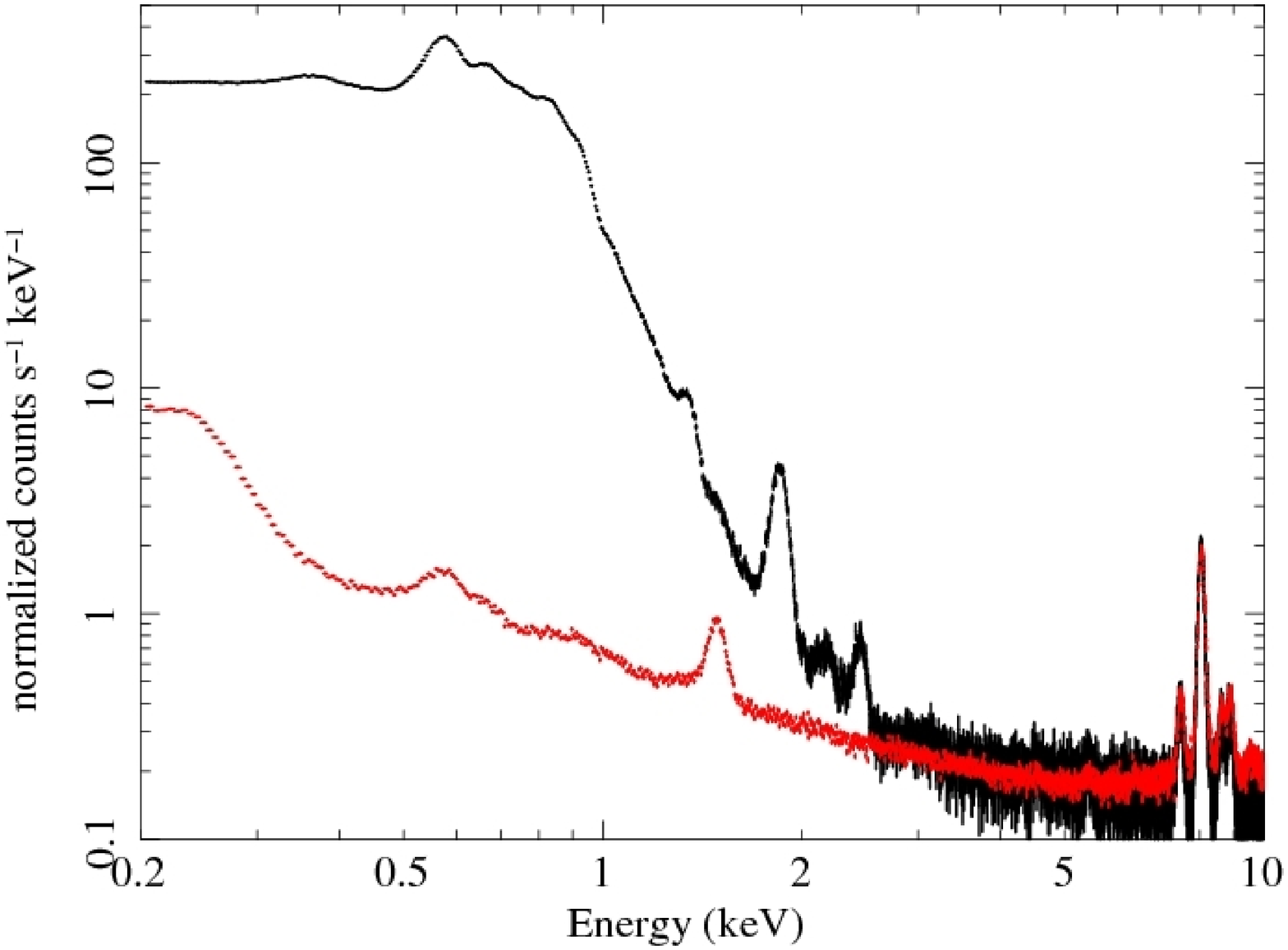}
  \end{center}
  \caption{Spectra from the entire FOV of the Cygnus Loop. Top, middle, and bottom panels show spectra obtained from XIS, MOS, and pn, respectively. Background spectra are also shown in red. }\label{fig:All_BGD}
\end{figure}

\section{Spectral Analysis}
\subsection{Line Features at 3.0 keV}
As shown in figure \ref{fig:HRI}, our FOV covers more than half the region of the Cygnus Loop.
We improved a signal-to-noise ratio and created a global spectrum of the Cygnus Loop by using all the available data listed in tables \ref{tab:sum_suzaku} and \ref{tab:sum_xmm}.
We should note that the exposure time is different from region to region and some parts have not yet been observed.
Therefore, the spectrum compiled in this way is not the precise one of the entire Cygnus Loop.

First, we created a spectrum from each observing region and combined them with FTOOLS \textit{mathpha}. 
In this analysis, we treated the XIS, MOS, and pn data independently.
In order to generate response matrix files (RMFs) and ancillary response files (ARFs), we employed \textit{xisrmfgen} and \textit{xissimarfgen} for the \textit{Suzaku} data \citep{Ishisaki07}, \textit{rmfgen} and \textit{arfgen} for the \textit{XMM-Newton} data.
We added the RMFs and the ARFs  with FTOOLS \textit{addrmf} and \textit{addarf}, respectively.
Thus, we obtained the combined XIS, MOS, and pn spectra. 
Figure \ref{fig:All_BGD} shows the spectra from the entire FOV of the Cygnus Loop. 
Left, middle, and right panels show spectra obtained from the XIS, MOS, and pn, respectively. 
We also show background spectra in red lines.
In these spectra, we can see many emission lines: C Ly$\alpha$, N He$\alpha$, O He$\alpha$, O Ly$\alpha$, the Fe-L complex, Ne He$\alpha$, Ne Ly$\alpha$, Mg He$\alpha$, Si He$\alpha$, Si He$\beta$, and S He$\alpha$.
Since these lines are all located below $\sim$3.0 keV, they are clearly detected above the background level.
It is difficult to detect line emissions above $\sim$3.0 keV because the X-ray emission is buried in the background level; the Cygnus Loop is a middle-aged SNR, and the electron temperature is much lower (0.1-0.9 keV; \cite{Uchida09a}) than those of the younger SNRs.
However, in figure \ref{fig:All_BGD}, some line structures are seen around 3.0 keV above the background level due to the improvement of the statistics.
These structures are likely S He$\beta$ and Ar He$\alpha$ lines which have not been detected by previous observations.
They are seen in all three spectra, and it is particularly prominent in the XIS spectrum due to the low background level of the \textit{Suzaku} satellite.

\begin{figure}
  \begin{center}
    \FigureFile(80mm,80mm){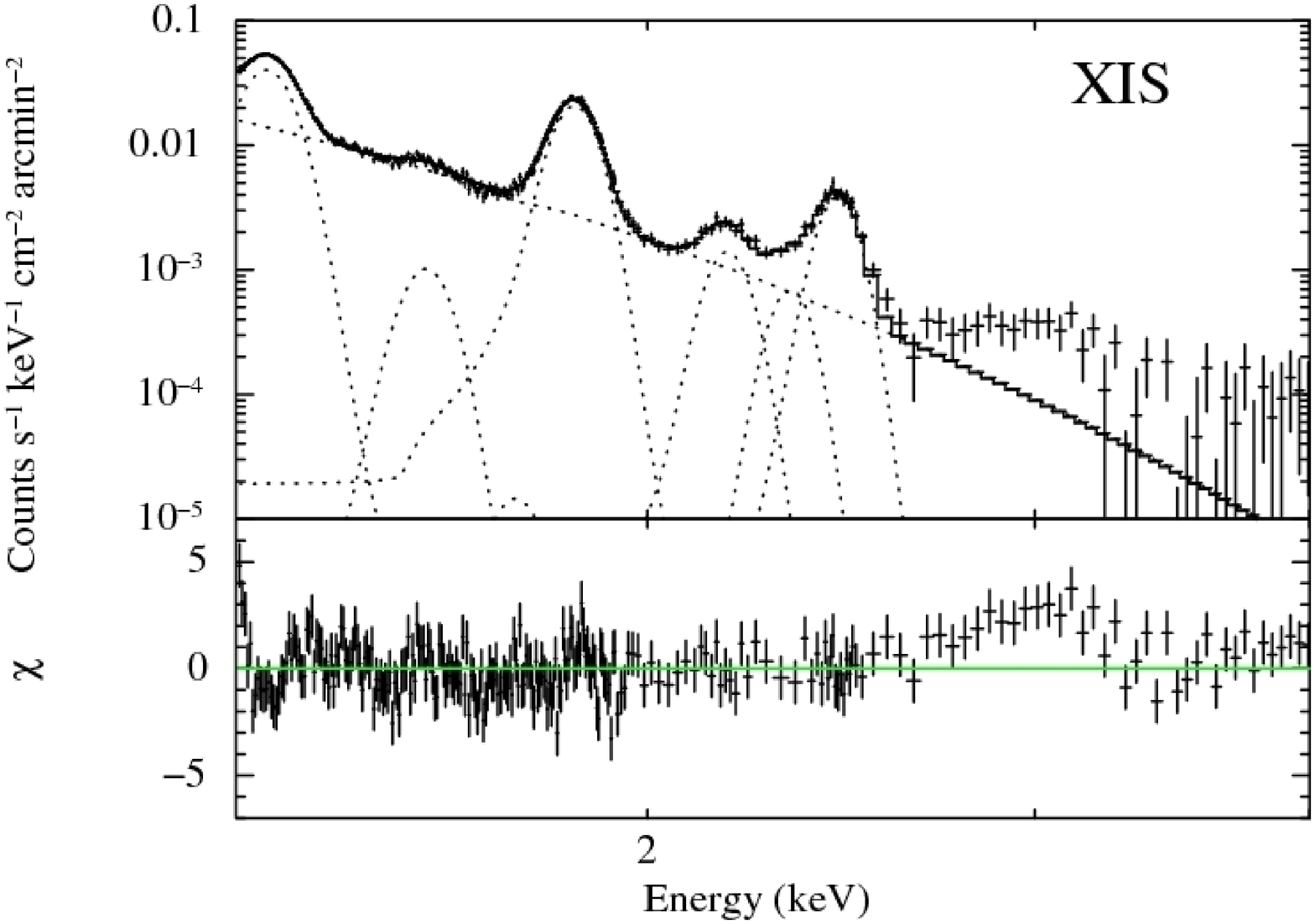}
    \FigureFile(80mm,80mm){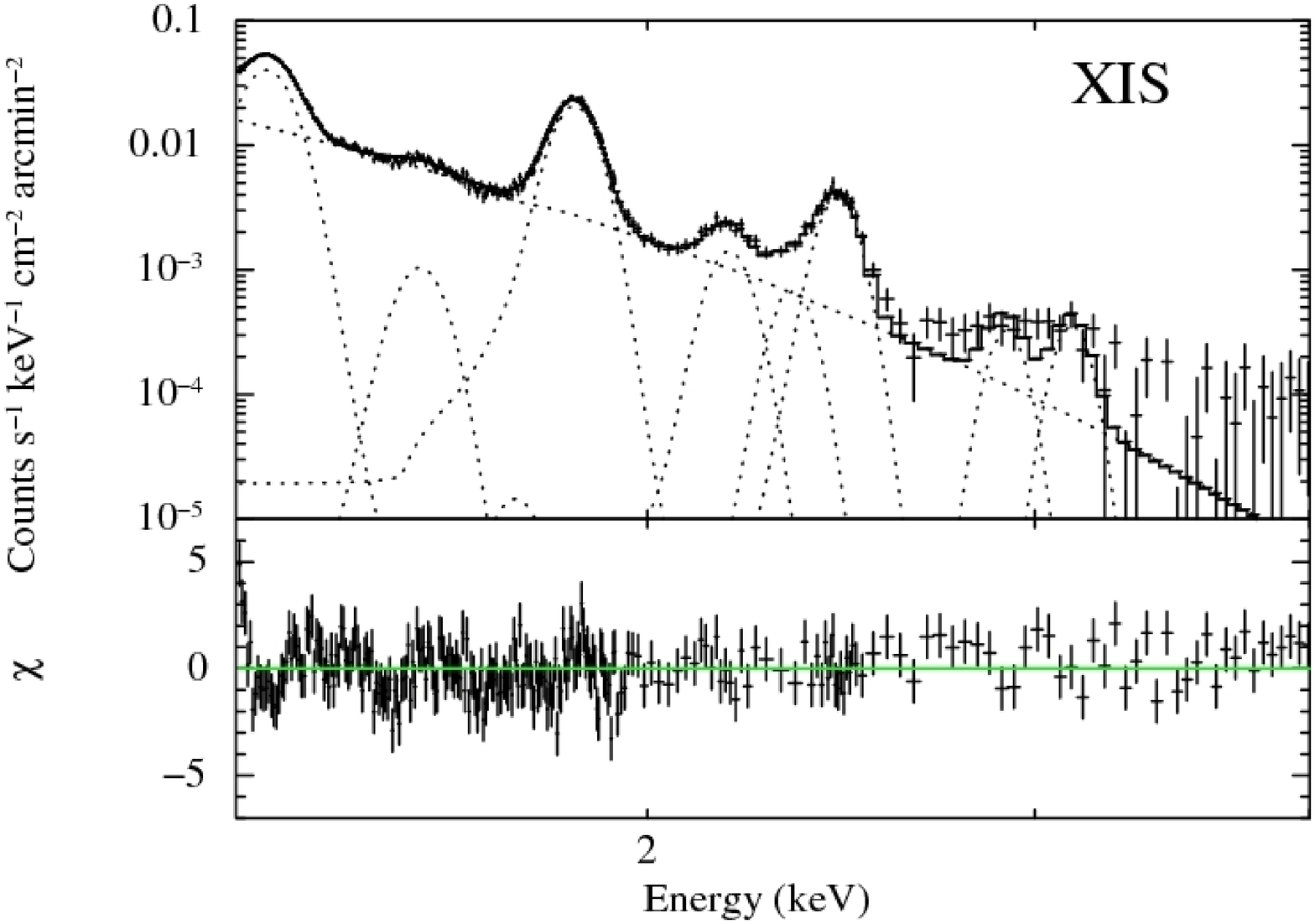}
    \FigureFile(80mm,80mm){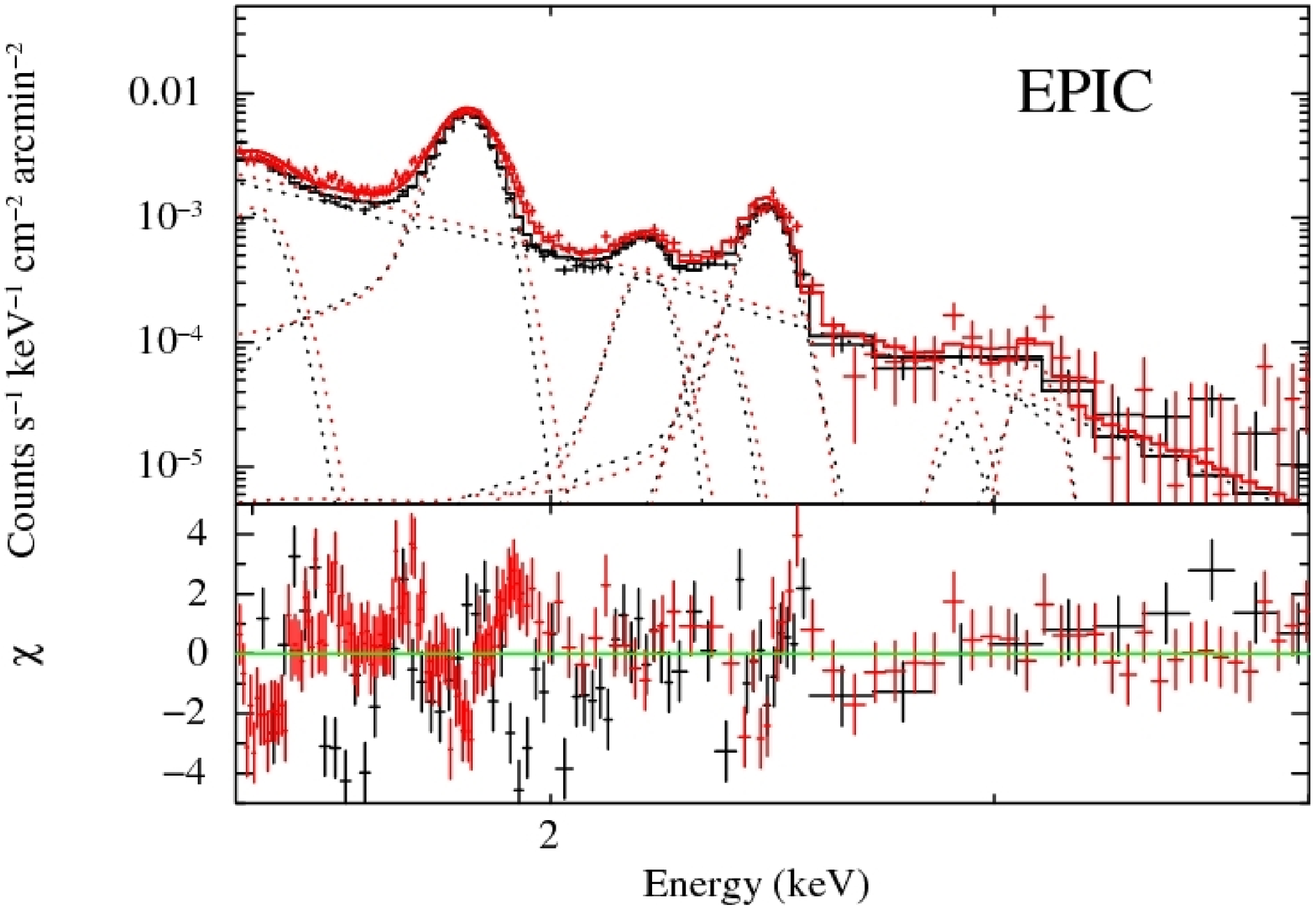}
  \end{center}
  \caption{The 1.5-5.0 keV spectrum of the Cygnus Loop with two model fits overlaid. We combined all XIS data shown in figure \ref{fig:HRI} top. They are fitted with bremsstrahlung plus five Gaussian components (left) and bremsstrahlung plus seven Gaussian components (right). The bottom figure shows the result of EPIC data. Red and black represent MOS and pn spectra, respectively. The residuals are shown in the lower panels. The best-fit parameters are shown in table \ref{tab:XIS_Gau}.}\label{fig:XIS_Gau}
\end{figure}

Figure \ref{fig:XIS_Gau} upper two panels show 1.5-5.0 keV spectra obtained from the XIS data with different models. 
First we fitted the spectrum with  bremsstrahlung plus five Gaussian components as shown in figure \ref{fig:XIS_Gau} left.
In this model, the Gaussian components correspond to the emission lines of Mg He$\beta$, Si He$\alpha$, Si He$\beta$, Si He$\gamma$, and S He$\alpha$.
The parameters of the electron temperature $kT_e$ and the emission measure (EM $= \int n_e n_{\rm H} dl$, where $n_e$ and $n_{\rm H}$ are the number densities of hydrogen and electrons and $dl$ is the plasma depth) in the bremsstrahlung model were left free. 
In the Gaussian models, we varied the line center energy and the normalization while we fixed the line width at zero.
We fixed the absorption column density $N\rm{_H}$ to be 2.3$\times10^{20}$cm$^{-2}$ that is an averaged value based on the previous analysis of the \textit{Suzaku} and the \textit{XMM-Newton} \citep{Uchida09b}.
The best-fit parameters are shown in table \ref{tab:XIS_Gau}.
In figure \ref{fig:XIS_Gau} left, the spectrum shows an excess from the model at about 3.0 keV.
Thus we added two more Gaussian components to the model (figure \ref{fig:XIS_Gau} right).
Accordingly the reduced $\chi^2$ value is improved  and the F-test shows that the former model is rejected for a significance level of 99\%.
Furthermore, the line center energies of the extra components are derived to be 2936$^{+49}_{-61}$ eV and 3128$\pm$39 eV, respectively. 
We also fitted the MOS and the pn spectra with the same model (figure \ref{fig:XIS_Gau} bottom). 
The result clearly shows an existence of two Gaussian components at 2920$^{+50}_{-55}$ eV and  3132$^{+26}_{-29}$ eV, respectively.
Since these values are statistically consistent with the expected values for S He$\beta$ (2884 eV) and Ar He$\alpha$ (3104 eV) lines according to the Astrophysical Plasma Emission Code (APEC; \cite{Smith01}), we conclude that  the excess structure around 3.0 keV is the S He$\beta$ and Ar He$\alpha$ lines.

\begin{figure}
  \begin{center}
    \FigureFile(80mm,80mm){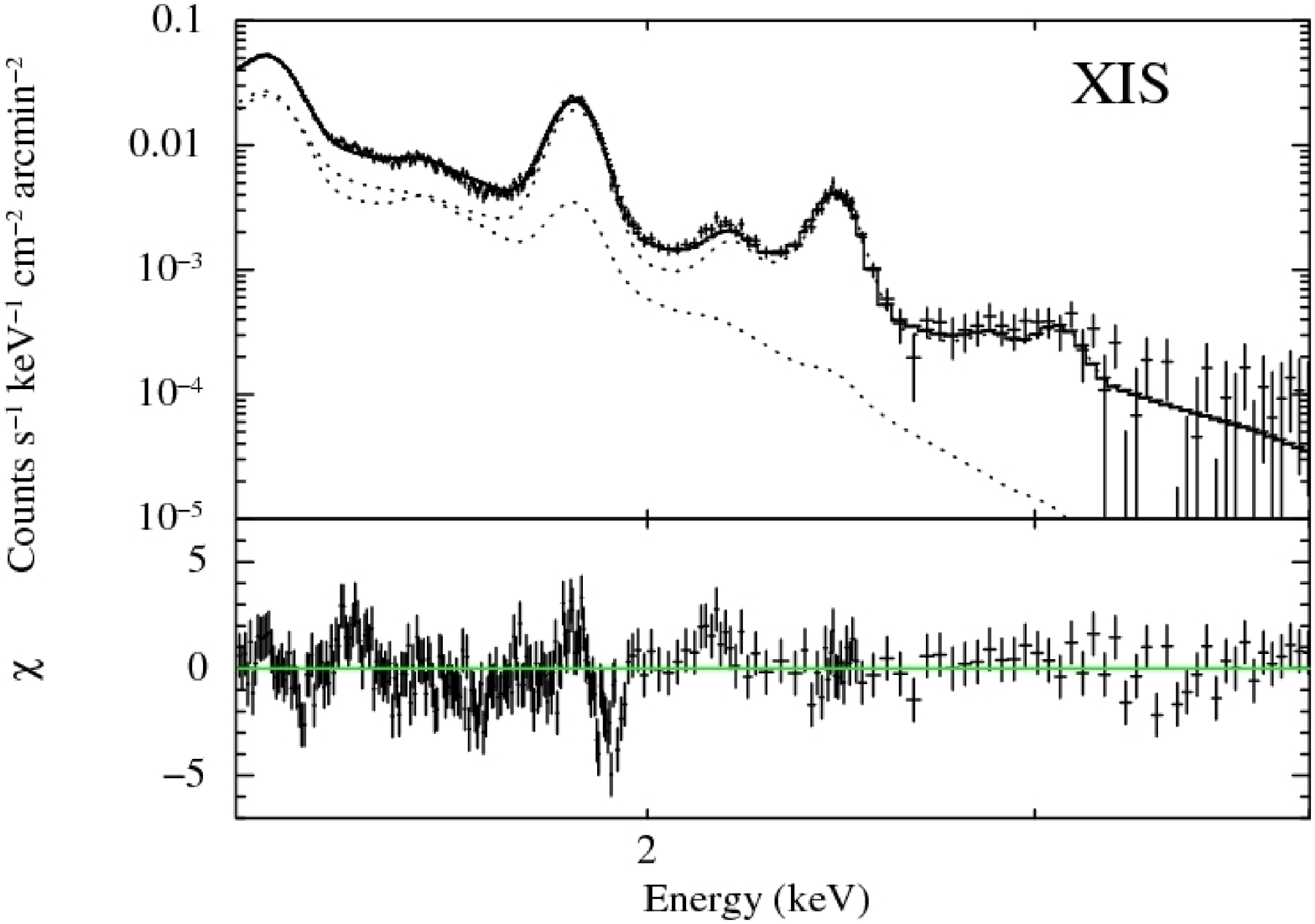}
    \FigureFile(80mm,80mm){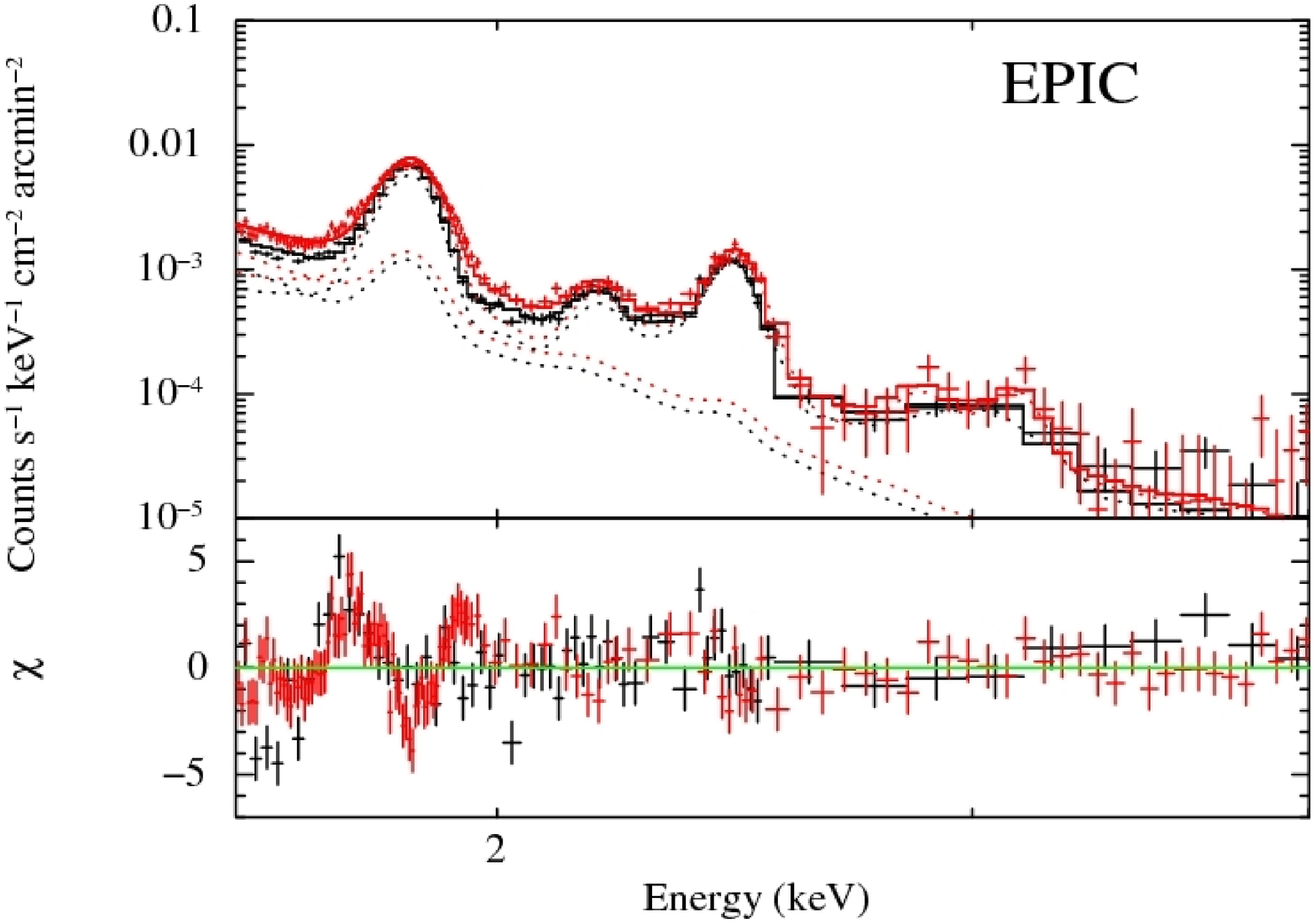}
  \end{center}
  \caption{\textit{Left}: Spectra obtained from all XIS data.  \textit{Right}: Spectra obtained from all EPIC MOS data (red) and EPIC pn data (black). Each spectrum is fitted with the two-component VPSHOCK model.  The residuals are shown in the lower panels. The best-fit parameters are shown in table \ref{tab:FIBIEPIC}.}\label{fig:FIBIEPIC}
\end{figure}

\subsection{Abundance Measurement of Ar}
The X-ray spectrum of the Cygnus Loop is known to be expressed by two components;
a high-temperature ejecta component and a low-temperature interstellar medium (ISM) component.
\citet{Tsunemi07} observed the Loop from northeast to southwest and reported that the high-$kT_e$ component varies between 0.4 keV and 0.9 keV.
\citet{Uchida09b} observed the rim of the Loop and showed that the temperature of the low-$kT_e$ component ranges from 0.12 keV to 0.35 keV.
They found that the temperature and the ionization timescale of each component are not uniform in the Loop.
In order to measure the Ar abundance, we basically follow the two-component model with the NEIver 2.0 spectral code that includes K-shell emission line of Ar, but we introduced an augmented NEIver 2.0 (ver 2.0$+$) model (c.f., \cite{Badenes06}; \cite{Park10}) which includes various inner-shell lines.
In addition, we prefer a VPSHOCK model (based on the NEIver 2.0$+$) to a simple VNEI because the VPSHOCK represents a more general case of the ionization timescale parameter than VNEI.
For the following analysis, we also set the offset of the gain to be free by using \textit{gain fit} in XSPEC for both detectors.
We note that the best-fit offset values are $\sim-$5 eV which is within the energy uncertainty of the XIS ($\pm$0.2$\%$ measured with the $^{55}$Fe calibration sources\footnote{The Suzaku Technical Description: http://www.astro.isas.ac.jp/suzaku/doc/suzaku\_td/suzaku\_td.html}) and EPIC (5 eV and 10 eV for MOS and pn, respectively\footnote{EPIC Calibration Status: http://xmm2.esac.esa.int/external/xmm\_sw\_cal/calib/index.shtml}).

Consequently, we applied a two-component VPSHOCK model (NEIver 2.0$+$):
In the low-$kT_e$ component, we fixed the values of $kT_e$ and upper limit of ionization timescale $\tau_{upper}$ (a product of the electron density and the elapsed time after the shock heating) to the averaged values based on the previous limb observations \citep{Uchida09b}.
We also fixed the metal abundances (in units of solar) to the values based on their results.
The values are summarized in table \ref{tab:shell}.
We fixed other elements (He, Ar, and Ca) to the solar values.
We varied the abundances of O, Ne, Mg, Si, S, Ar, and Fe in the high-$kT_e$ component, and set those of  C and N equal to O, Ni equal to Fe.
The other elements (He, and Ca) were fixed to their solar values.
Other parameters such as $kT_e$ and $\tau_{upper}$ were all free.
We also set the absorption column density, $N\rm{_H}$, to be free.

As a preliminary step, we applied  the two-component model for every full-band (0.3-5.0 keV) spectrum obtained from each FOV.
Although this model fitted the spectra well, we could not constrain the abundance of Ar.
Therefore, we next applied the same model to the spatially-integrated spectra in order to measure the Ar abundance. 
We then found that the fits are far from acceptable from a statistical point of view.
The most noticeable discrepancies between the data and the best-fit model were found around 0.8 keV where the X-ray spectrum in figure \ref{fig:All_BGD} shows the highest count rate (which is about 1000 times higher than that around 3 keV).
As mentioned above, the low energy band ($\lesssim$1 keV) is dominated by the low temperature component, whereas the high energy band is dominated by the high temperature component. 
It is almost certain that the Ar lines come from the high temperature component and that the low energy band are not important to measure the Ar abundance.
Thus, we here focus on the X-ray energy range above 1.3 keV.
The discrepancies found at the low energy band is very interesting.
However, detailed investigations are beyond the scope of this paper, and are left as our future work.
As for the EPIC data, an excess exists around 1.5 keV which corresponds to the neutral Al-K line emission due to an instrumental fluorescence.
Thus we used the data above 1.6 keV for EPIC data.
In these fits, we fixed the abundances for O, Ne, Mg (only for EPIC spectrum), and Fe to the averaged values of the results calculated from every FOV.

\begin{figure}
  \begin{center}
    \FigureFile(90mm,90mm){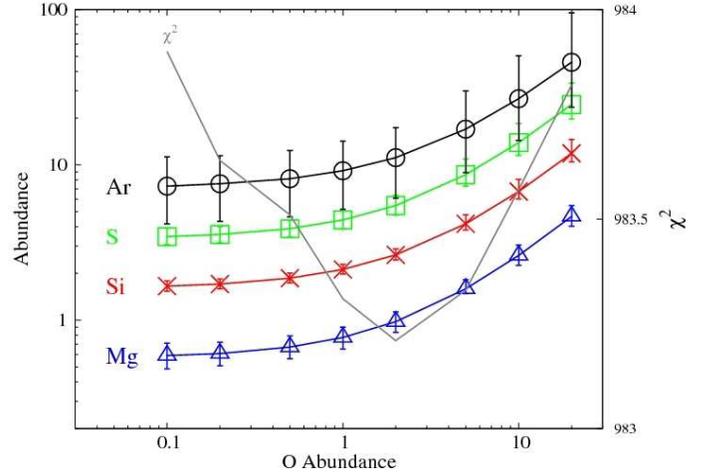}
 \end{center}
  \caption{Best-fit parameters of Mg (triangle), Si (X), S (square), and Ar (circle) with the two-component VPSHOCK model are plotted as a function of O abundance. The values of $\chi^2$ are overlaid.}\label{fig:Ofix}
\end{figure}

Figure \ref{fig:FIBIEPIC} left shows the fitting result of the XIS spectrum. 
The best-fit parameters are shown in table \ref{tab:FIBIEPIC}.
The spectrum is well fitted with the two-component VPSHOCK model ($\chi^2$/dof=1096/1004).
By comparing with several previous observations (e.g., \cite{Tsunemi07}), it is reasonable to consider that the temperature of the high-$kT_e$ component, 0.69$^{+0.04}_{-0.02}$ keV, is an averaged one of the Cygnus Loop.  
The values of Mg, Si and S abundances are also consistent with the previous studies.
From the results, it is clear that the Ar abundance of the Cygnus Loop is significantly higher than the solar value (9.0$^{+4.0}_{-3.8}$).
We note that the calculated Ar abundance may significantly depend on the fixed values of O (and the other) abundances.
Then, in order to check the effect of the fixed parameters on the abundance measurement, we varied the values of O, Ne, and Fe abundances of the high-$kT_e$ component from 0.1 to 20 and calculated each best-fit parameter.
Figure \ref{fig:Ofix} shows the best-fit parameters of Mg, Si, S, and Ar abundances for various values as a function of O, Ne, and Fe abundance.
We also overlaid the $\chi^2$ value on it.
We found that the metal abundances depend on the O abundance.
However, we found the abundances for Ar, Si and S are above the solar value.
In particular, the Ar shows the highest abundance among them.
Therefore, we concluded that the Ar abundance of the Cygnus Loop is significantly higher than the solar value.
We also confirmed this result from the EPIC spectrum (see figure \ref{fig:FIBIEPIC} right and table \ref{tab:FIBIEPIC}).
The EPIC result also shows a significant high abundance of Ar (8.4$^{+2.5}_{-2.7}$) that is consistent with the XIS result within a statistical error.
Thus, we concluded that the Ar He$\alpha$ line is detected from the Cygnus Loop's ejecta for the first time.

\section{Discussion}
We investigate the spatial distribution of Ar using the \textit{Suzaku} XIS data.
Due to the lack of statistics, we are not able to subdivide the \textit{XMM-Newton} data.
We divide the XIS FOV into three regions as shown in figure \ref{fig:HRI}.
The regions surrounded by red, blue, and black lines are labeled as Region-A, Region-B, and Region-C, respectively.

\begin{figure}
  \begin{center}
    \FigureFile(80mm,80mm){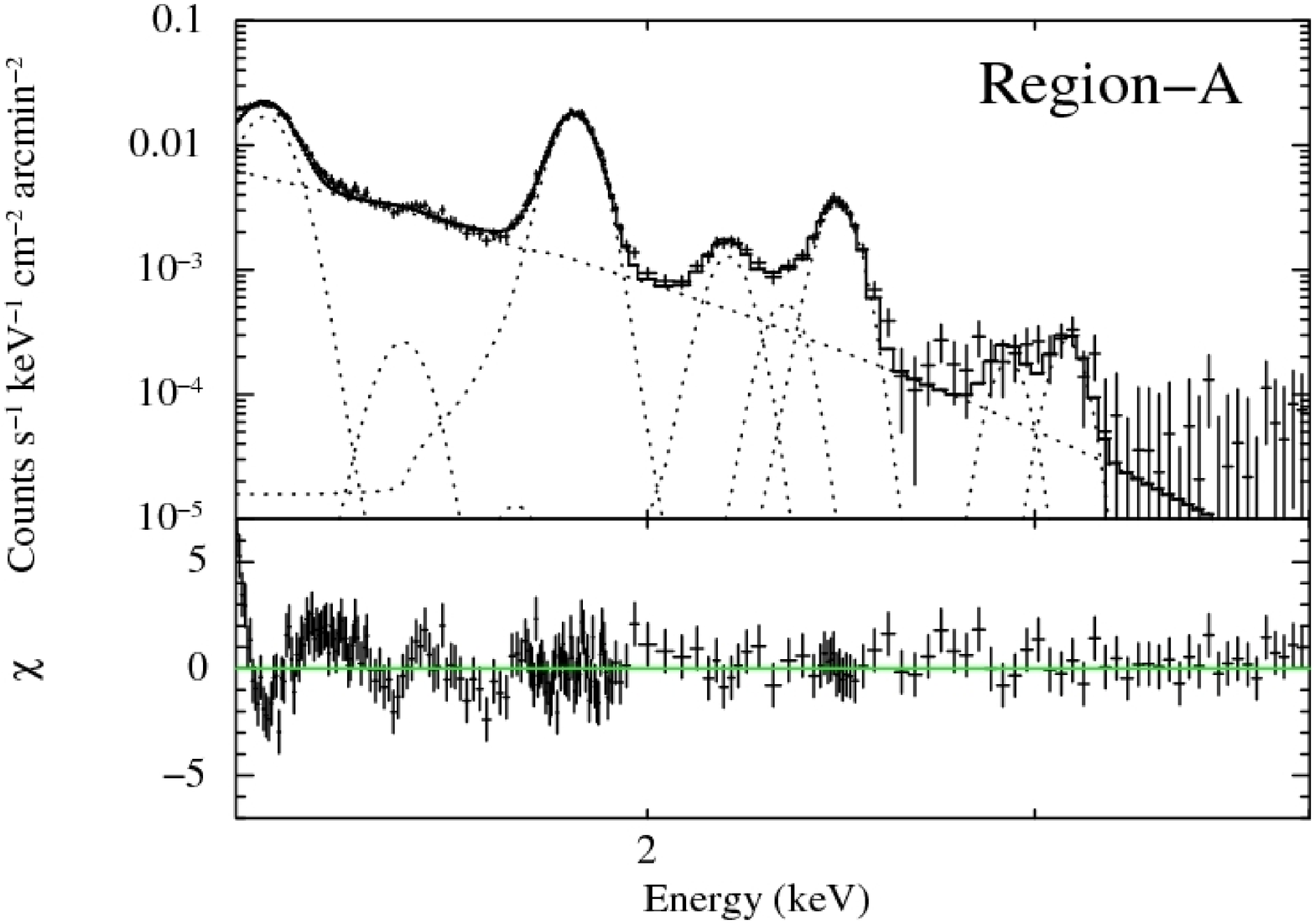}
    \FigureFile(80mm,80mm){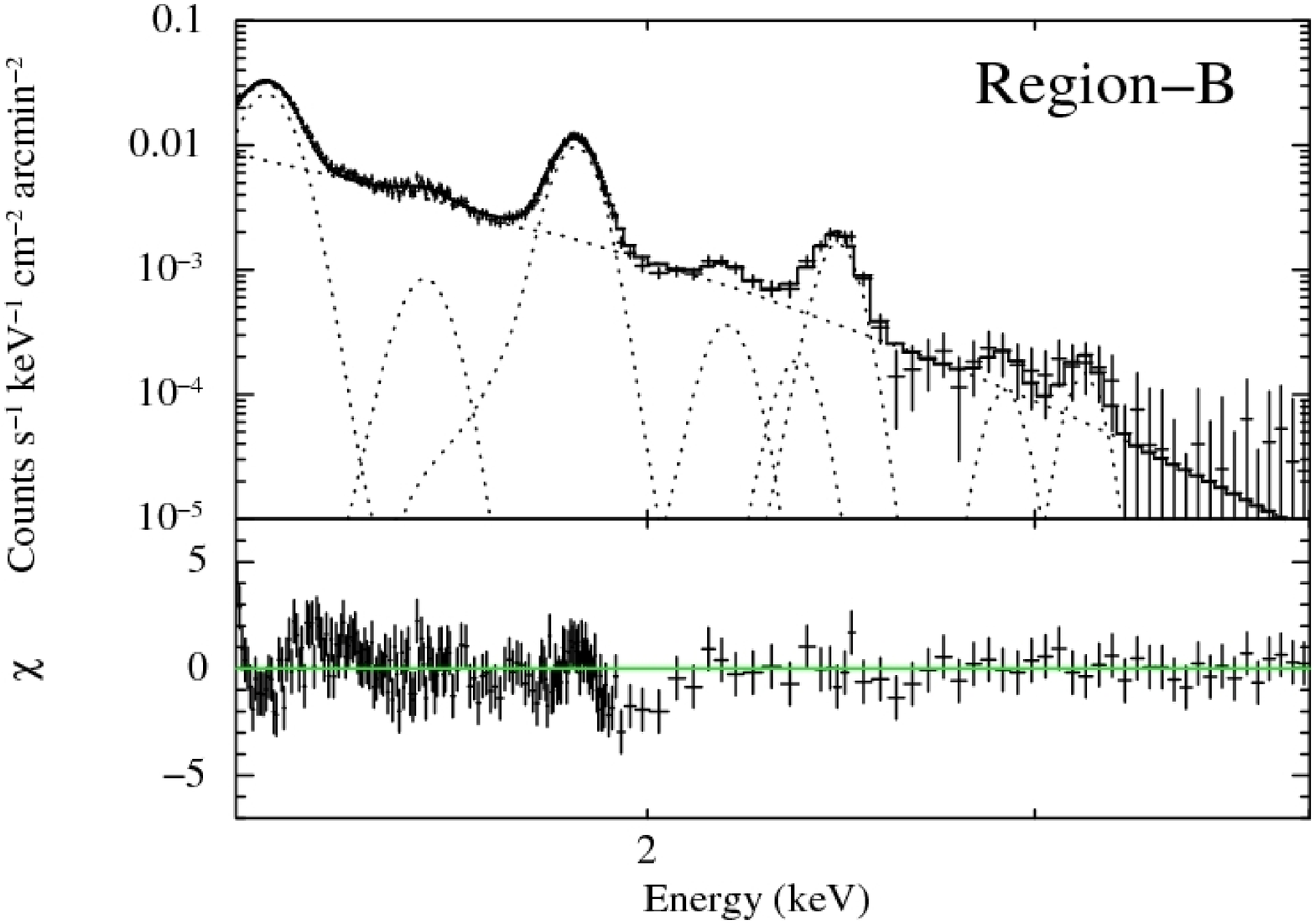}
    \FigureFile(80mm,80mm){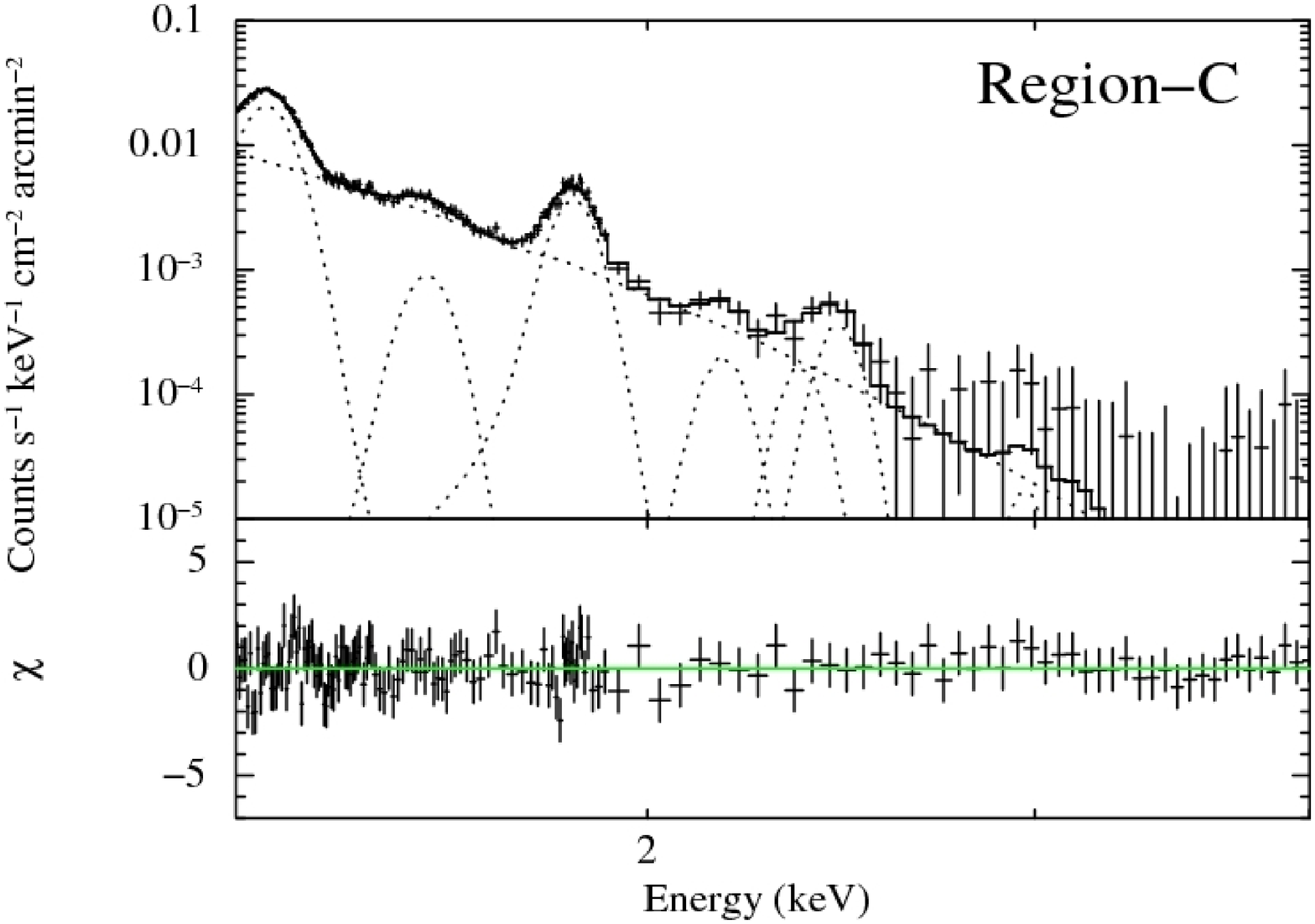}
  \end{center}
  \caption{The 1.3-5.0 keV spectra of the Cygnus Loop. Spectral extraction regions are shown in figure \ref{fig:HRI} top. They are fitted with bremsstrahlung plus seven Gaussian components. The residuals are shown in the lower panels. The best-fit parameters are shown in table \ref{tab:Region_Gau}.}\label{fig:Region_Gau}
\end{figure}

\begin{figure}
  \begin{center}
     \FigureFile(80mm,80mm){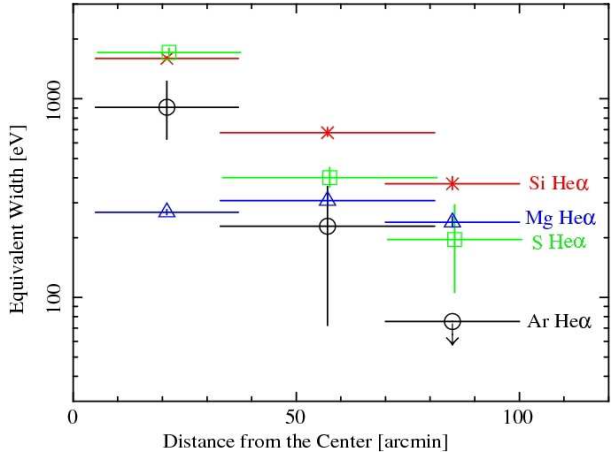}
  \end{center}
  \caption{Equivalent width distributions of Mg He$\beta$ (triangle), Si He$\alpha$ (X), S He$\alpha$ (square), and Ar He$\alpha$ (circle).  The horizontal bars in the x-axis represent the range of each region. The upper limit is obtained from the result of Region-C. }\label{fig:eqwidth}
\end{figure}

Firstly, we applied the bremsstrahlung plus seven Gaussian components for each spectrum.
The fitting method is the same as the one explained in the previous section. 
The results and the best-fit parameters are shown in figure \ref{fig:Region_Gau} and table \ref{tab:Region_Gau}, respectively. 
From figure \ref{fig:Region_Gau}, the Gaussian-like structures are seen at about 3.0 keV particularly in the Region-A spectrum.
We note that the line-center energy  corresponds to that of the Ar He$\alpha$ line for each fit.
The contribution of Ar He$\alpha$ line emission is decreasing from Region-A to Region-C, i.e., from the Cygnus Loop center to the rim, suggesting the observed Ar is concentrated on the center of the Loop.
Figure \ref{fig:eqwidth} shows the radial distributions of the equivalent widths for some lines (Mg He$\beta$, Si He$\alpha$, S He$\alpha$, and Ar He$\alpha$).
From figure \ref{fig:eqwidth}, the equivalent widths of Si and S lines are clearly decreasing from the center to the rim.
On the other hand, the equivalent width of Mg line is almost constant in our FOV.
According to the previous study (e.g., \cite{Uchida09a}), the Si and the S ejecta are concentrated on the Cygnus Loop's center, while the Mg ejecta are almost uniform from the center to the rim.
Therefore, the equivalent-width distributions of Si, S and Mg reflect their radial profiles of the ejecta. 
While the equivalent widths of the Ar line have relatively large errors, they are also decreasing toward the rim as well as those of the Si and S distributions.
This result strongly suggests that the Ar emission is also derived from the progenitor star.

\begin{figure}
  \begin{center}
    \FigureFile(80mm,80mm){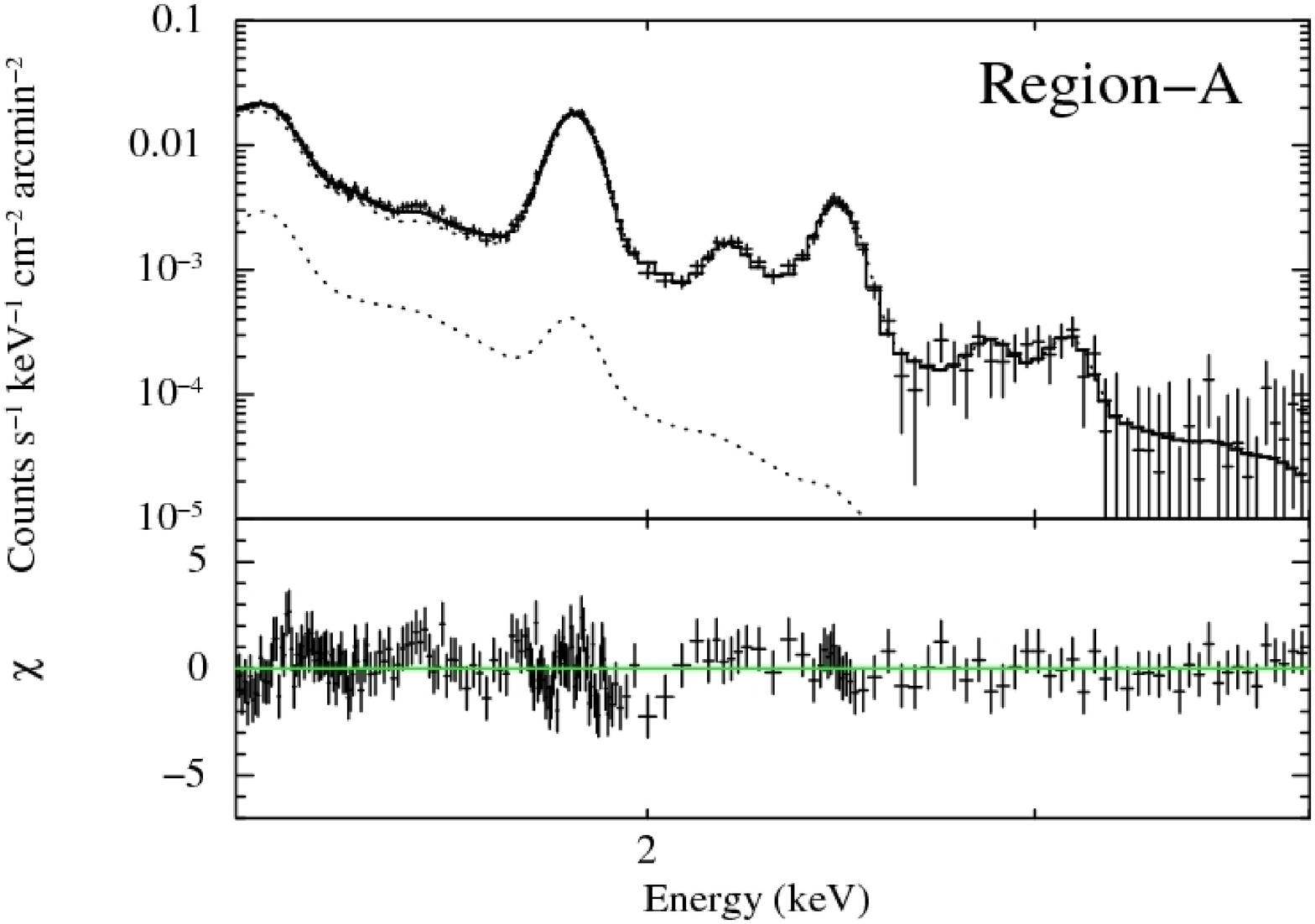}
    \FigureFile(80mm,80mm){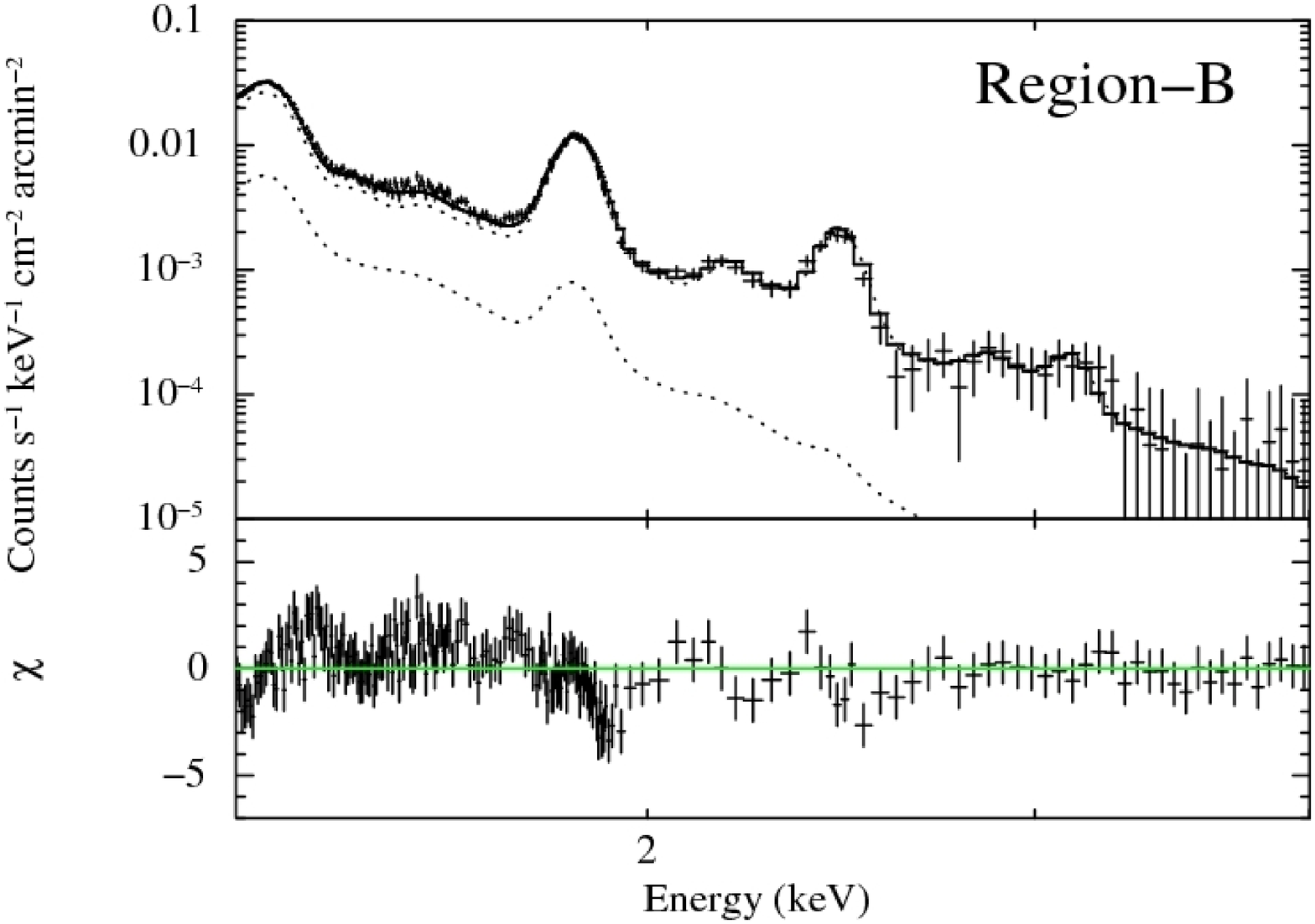}
    \FigureFile(80mm,80mm){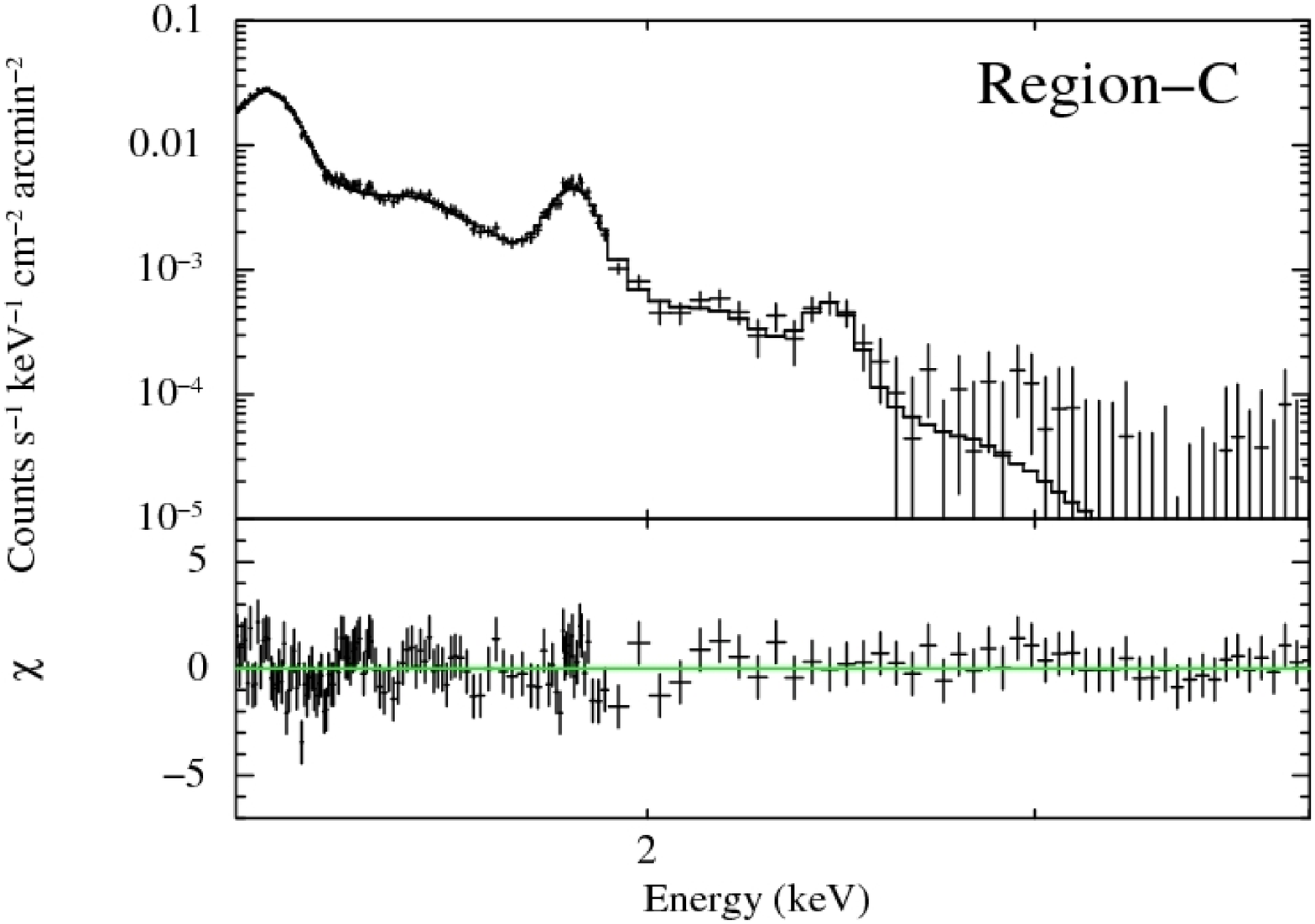}
  \end{center}
  \caption{The 1.3-5.0 keV spectra of the Cygnus Loop with two-component VPSHOCK model (top panels) and one-component VPSHOCK model (bottom panel). The residuals are shown in the lower panels. The best-fit parameters are shown in table \ref{tab:Region_NEI}.}\label{fig:Region_NEI}
\end{figure}

In order to confirm it, we next measured the Ar abundance of each region by applying the two-component VPSHOCK model as described in the previous section.
The spectra from Region-A and Region-B are the superposition of both ISM and ejecta components.
On the other hand, the spectrum from Region-C is virtually dominated by the emission from the ISM component.
Therefore, we applied one-component VPSHOCK model for the Region-C spectrum.
Figure \ref{fig:Region_NEI} shows the spectra obtained from Region-A (top), Region-B (middle), and Region-C (bottom). 
The best-fit parameters are indicated in table \ref{tab:Region_NEI}.
Each $\chi^2$ value shows that two/one-component VPSHOCK models are statistically acceptable for Region-A/B and Region-C, respectively.
From table \ref{tab:Region_NEI}, both the spectra from Region-A and Region-B show that the Ar abundances are significantly higher than that of the solar value.
Furthermore, the best-fit value of the Ar abundance is decreasing toward the rim as well as those of Si and S.
On the other hand, the distribution of Mg abundance is almost flat.
A previous spatially-resolved analysis showed the similar distributions of Si and Mg \citep{Uchida09a}.
They concluded that these elements are derived from the ejecta blown off by a core-collapse SN explosion.
From table \ref{tab:Region_NEI}, the co-existence of Si, S and Ar is consistent with a theoretical picture that Si, S, and Ar are mainly produced in the static O-burning layer during the stellar evolution and the incomplete Si-burning layer during the SN explosion.
Thus we confirmed that the observed Ar emission is originated from the ejecta of the Cygnus Loop rather than the ISM. 
Due to the lack of the statistics, we were not able to subdivide our FOV to investigate the spatial distribution of Ar in more detail.
However, considering the co-existence of Si and Ar abundances, the Ar-rich ejecta may also have a central concentration similar to that of Si.

\begin{figure}
  \begin{center}
    \FigureFile(80mm,80mm){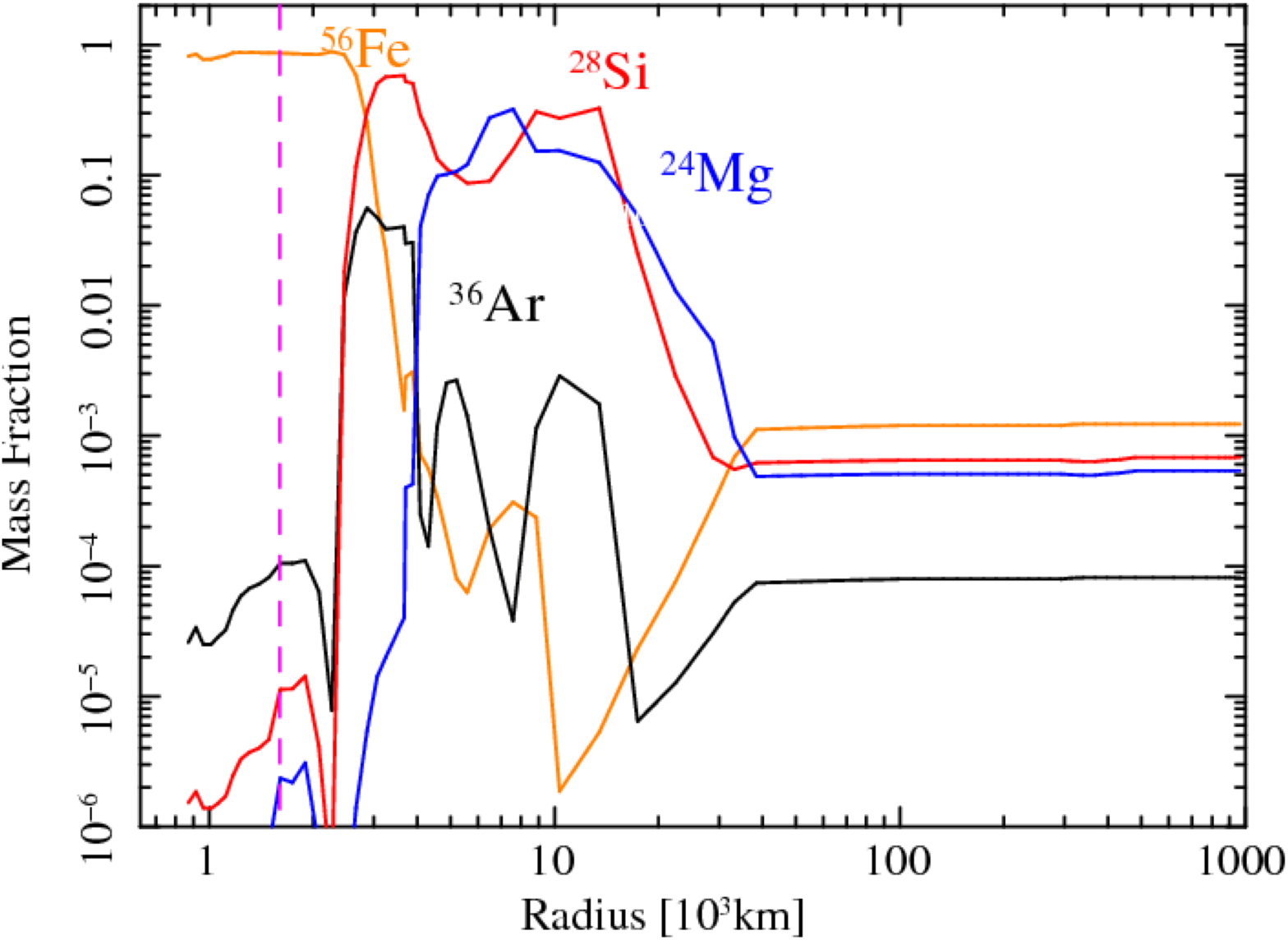}
  \end{center}
  \caption{Mass fractions of $^{24}$Mg, $^{28}$Si, $^{36}$Ar, and $^{56}$Fe in a 15\MO  \ star after taking into account the radioactive decays (based on the results of \cite{Umeda05, Nomoto06, Tominaga07}). The dotted vertical line represents a boundary between the ejecta and the central remnant of 1.50\MO. 
}\label{fig:FeOAr}
\end{figure}

Some previous studies estimated the progenitor mass of the Cygnus Loop to be $\sim$15\MO \ (\cite{Levenson98}; \cite{Tsunemi07}), hence the core-collapse origin.
However, our results indicate that the O, Ne, and Mg abundances are significantly lower than those of Si, S, and Ar, which is inconsistent with the standard theoretical core-collapse model.
Although there may exist O-Ne-Mg-rich region outside our FOV as suggested by \citet{Uchida09a}, the low abundances of O-Ne-Mg are still remains an open question.
Since Ar is produced nearer the center of the Fe core than the other lighter elements, the abundance measurement will be useful for the future estimation of the progenitor mass by expanding the observing region and improving the statistics.
Figure \ref{fig:FeOAr} shows an example mass fractions  of $^{24}$Mg, $^{28}$Si, $^{36}$Ar, and $^{56}$Fe in a 15\MO \ solar-metallicity star as a function of the presupernova radius which takes into account explosive nucleosynthesis and radioactive decay \citep{Umeda05, Nomoto06, Tominaga07}.
The dotted vertical line represents a boundary between the ejecta and the central remnant to yield a canonical amount of $^{56}$Ni ($\sim$0.08\MO; e.g., SN1987A; \cite{Blinnikov00}), which locates at $R=1600$ km in the progenitor star corresponding to 1.50\MO.
Since the Ar-rich layer exists near the boundary ($R>2080$ km), the mass of the ejected Ar highly depends on the position of the boundary as well as that of Fe.
For example, if the mass of the central remnant increases only 2\% (i.e., 1.53\MO), the boundary reaches the Ar-rich layer, which may have a large influence on the abundance of the ejected Ar.
While Si and S are also produced near the boundary, these elements are abundant as well even in the O-Ne-Mg layer (see figure \ref{fig:FeOAr}).
Consequently, the total amount of the ejected Ar more significantly depends on the boundary, i.e., progenitor mass than those of Si and S.
In general, the Fe abundance provides us with an important clue to exploring the type of the SN explosion.
However, as for the Cygnus Loop, the low plasma temperature makes it difficult to detect the Fe-K line emissions in X-ray.
Therefore, the accurate measurement of Ar abundance will be important in determining the mass of the progenitor star and the central compact object that still has not been discovered.

\section{Summary}
In summary, we detected Ar He$\alpha$ and S He$\beta$ line emissions from the Cygnus Loop for the first time by accumulating all the available data of \textit{Suzaku} and \textit{XMM-Newton}.
The Ar abundances obtained from the XIS and EPIC spectra are 9.0$^{+4.0}_{-3.8}$ and 8.4$^{+2.5}_{-2.7}$, respectively.
The high metal abundances clearly show that the Ar line originates from the ejecta of the Cygnus Loop.
We also found that the central concentration  of Ar is closer to the distribution of Si and S rather than that of Mg. 
The measurement of Ar abundance will be important in determining the progenitor mass of the Loop for the future analysis.


\section*{Acknowledgments}
H.U. thanks Professor Jacco Vink and his students for many useful discussions and their hospitality at Utrecht University.
The authors would like to thank H. Umeda for providing a progenitor model.
H.U. and M.K. are supported by JSPS Research Fellowship for Young Scientists. 
S.K. is supported by JSPS Postdoctoral Fellow for Research Abroad.

\onecolumn

\begin{table}
\caption{Summary of the 32 \textit{Suzaku} observations\label{tab:sum_suzaku}}
\begin{center}
\begin{tabular}{lccc}
\hline
\hline
\multicolumn{4}{c}{\textit{Suzaku} Observations} \\
\hline
Obs. ID  & Obs. Date & R.A., Dec. (J2000) & Effective Exposure\\
\hline
	   						    		      			  						
500020010   & 2005-Nov-23 &  20$^{\mathrm h}$56$^{\mathrm m}$48$^{\mathrm s}$.9, 31\arcdeg56\arcmin54\arcsec.8  & 20.4 ks\\
	   					    		      			  		    			
500021010   & 2005-Nov-24 &  20$^{\mathrm h}$55$^{\mathrm m}$56$^{\mathrm s}$.0, 31\arcdeg56\arcmin53\arcsec.2  & 21.4 ks\\
	   					    		      			  		    			
500022010   & 2005-Nov-29 &  20$^{\mathrm h}$55$^{\mathrm m}$05$^{\mathrm s}$.6, 32\arcdeg10\arcmin35\arcsec.4  & 21.7 ks\\
							    		      			  		    			
500023010   & 2005-Nov-30 &  20$^{\mathrm h}$54$^{\mathrm m}$03$^{\mathrm s}$.8, 32\arcdeg21\arcmin47\arcsec.9  & 25.3 ks\\
	   						    						    		    
501029010   & 2006-May-09 &  20$^{\mathrm h}$55$^{\mathrm m}$00$^{\mathrm s}$.0, 31\arcdeg15\arcmin46\arcsec.8  & 13.2 ks\\
	   						    						    		    
501030010   & 2006-May-10 &  20$^{\mathrm h}$53$^{\mathrm m}$59$^{\mathrm s}$.3, 31\arcdeg03\arcmin39\arcsec.6  & 13.9 ks\\
	   						    						    		    
501031010   & 2006-May-12 &  20$^{\mathrm h}$52$^{\mathrm m}$58$^{\mathrm s}$.8, 30\arcdeg51\arcmin32\arcsec.4  & 18.2 ks\\
	   							   						  		   
501028010   & 2006-May-13 &  20$^{\mathrm h}$55$^{\mathrm m}$56$^{\mathrm s}$.3, 31\arcdeg28\arcmin56\arcsec.2  & 4.9 ks\\
	   						    						    		    
501033010   & 2006-May-22 &  20$^{\mathrm h}$50$^{\mathrm m}$58$^{\mathrm s}$.8, 30\arcdeg27\arcmin00\arcsec.0  & 20.0 ks\\

501034010   & 2006-May-22 &  20$^{\mathrm h}$48$^{\mathrm m}$49$^{\mathrm s}$.7, 30\arcdeg00\arcmin21\arcsec.6  & 13.9 ks\\
	   						    						    		    
501032010   & 2006-May-25 &  20$^{\mathrm h}$51$^{\mathrm m}$58$^{\mathrm s}$.6, 30\arcdeg39\arcmin10\arcsec.8  & 17.4 ks\\

501035010   & 2006-Dec-18 &  20$^{\mathrm h}$48$^{\mathrm m}$16$^{\mathrm s}$.2, 29\arcdeg42\arcmin07\arcsec.2  & 11.2 ks\\

501036010   & 2006-Dec-18 &  20$^{\mathrm h}$47$^{\mathrm m}$17$^{\mathrm s}$.3, 30\arcdeg04\arcmin21\arcsec.4  & 11.8 ks\\
	   							   						  		   
501017010   & 2007-Nov-11 &  20$^{\mathrm h}$49$^{\mathrm m}$11$^{\mathrm s}$.3, 30\arcdeg59\arcmin27\arcsec.6 & 28.7 ks\\
	   							   						  		   
501018010   & 2007-Nov-12 &  20$^{\mathrm h}$48$^{\mathrm m}$18$^{\mathrm s}$.7, 30\arcdeg46\arcmin33\arcsec.6 & 21.0 ks\\
	   							   						  		   
501019010   & 2007-Nov-12 &  20$^{\mathrm h}$47$^{\mathrm m}$14$^{\mathrm s}$.2, 30\arcdeg36\arcmin10\arcsec.8 & 16.2 ks\\
								   						  		   
501020010   & 2007-Nov-13 &  20$^{\mathrm h}$46$^{\mathrm m}$20$^{\mathrm s}$.8, 30\arcdeg23\arcmin22\arcsec.6 & 14.7 ks\\

501012010   & 2007-Nov-14 &  20$^{\mathrm h}$54$^{\mathrm m}$07$^{\mathrm s}$.6, 31\arcdeg57\arcmin22\arcsec.0 & 9.8 ks\\
	   							   						  		   
501013010   & 2007-Nov-14 &  20$^{\mathrm h}$53$^{\mathrm m}$08$^{\mathrm s}$.5, 31\arcdeg45\arcmin40\arcsec.3 & 16.4 ks\\
	   							   						  		   
501014010   & 2007-Nov-14 &  20$^{\mathrm h}$52$^{\mathrm m}$09$^{\mathrm s}$.9, 31\arcdeg36\arcmin43\arcsec.4 & 16.9 ks\\
	   							   						  		   
501015010   & 2007-Nov-14 &  20$^{\mathrm h}$51$^{\mathrm m}$11$^{\mathrm s}$.8, 31\arcdeg22\arcmin08\arcsec.4 & 18.3 ks\\
	   							   						  		   
501016010   & 2007-Nov-15 &  20$^{\mathrm h}$50$^{\mathrm m}$11$^{\mathrm s}$.3, 31\arcdeg10\arcmin48\arcsec.0 & 19.3 ks\\

503055010   & 2008-May-09 &  20$^{\mathrm h}$49$^{\mathrm m}$48$^{\mathrm s}$.7, 31\arcdeg30\arcmin18\arcsec.0  & 22.2 ks\\

503056010   & 2008-May-10 &  20$^{\mathrm h}$48$^{\mathrm m}$00$^{\mathrm s}$.0, 31\arcdeg10\arcmin30\arcsec.0  & 22.5 ks\\
	   					    		      			  		    			
503062010   & 2008-May-13 &  20$^{\mathrm h}$56$^{\mathrm m}$26$^{\mathrm s}$.5, 30\arcdeg19\arcmin55\arcsec.2  & 16.9 ks\\
	   						    		      			  	    			
503063010   & 2008-May-13 &  20$^{\mathrm h}$55$^{\mathrm m}$16$^{\mathrm s}$.3, 30\arcdeg01\arcmin44\arcsec.0  & 22.8 ks\\
	   					    		      			  		    			
503064010   & 2008-May-14 &  20$^{\mathrm h}$53$^{\mathrm m}$51$^{\mathrm s}$.6, 29\arcdeg54\arcmin42\arcsec.5  & 18.2 ks\\
	   						    						    		    
503057010   & 2008-Jun-02 &  20$^{\mathrm h}$52$^{\mathrm m}$43$^{\mathrm s}$.8, 32\arcdeg26\arcmin19\arcsec.0  & 16.2 ks\\
	   						    						    		    
503058010   & 2008-Jun-03 &  20$^{\mathrm h}$51$^{\mathrm m}$17$^{\mathrm s}$.2, 32\arcdeg25\arcmin24\arcsec.6  & 19.3 ks\\
	   						    						    		    
503059010   & 2008-Jun-03 &  20$^{\mathrm h}$49$^{\mathrm m}$50$^{\mathrm s}$.6, 32\arcdeg21\arcmin50\arcsec.8  & 19.5 ks\\
	   						    						    		    
503060010   & 2008-Jun-04 &  20$^{\mathrm h}$48$^{\mathrm m}$28$^{\mathrm s}$.2, 32\arcdeg17\arcmin44\arcsec.5  & 18.5 ks\\
	   					    		      			  		    			
503061010   & 2008-Jun-04 &  20$^{\mathrm h}$47$^{\mathrm m}$22$^{\mathrm s}$.7, 32\arcdeg10\arcmin22\arcsec.8  & 26.0 ks\\
\hline
\end{tabular}
\end{center}
\end{table}

\begin{table}
\caption{Summary of the 9 \textit{XMM} observations\label{tab:sum_xmm}}
\begin{center}
\begin{tabular}{lccc}
\hline
\hline
\multicolumn{4}{c}{\textit{XMM-Newton} Observations} \\
\hline
Obs. ID & Obs. Date & R.A., Dec. (J2000) & Effective Exposure\\
\hline

0082540101  & 2002-Nov-25 & 20$^{\mathrm h}$55$^{\mathrm m}$23$^{\mathrm s}$.6, 31\arcdeg46\arcmin17\arcsec.0  & 14.7 ks\\

0082540201  & 2002-Dec-03 & 20$^{\mathrm h}$54$^{\mathrm m}$07$^{\mathrm s}$.2, 31\arcdeg30\arcmin51\arcsec.1  & 14.4 ks\\
	   						    						     		    
0082540301  & 2002-Dec-05 & 20$^{\mathrm h}$52$^{\mathrm m}$51$^{\mathrm s}$.1, 31\arcdeg15\arcmin25\arcsec.7  & 11.6 ks\\
	   						    						     		    
0082540401  & 2002-Dec-07 & 20$^{\mathrm h}$51$^{\mathrm m}$34$^{\mathrm s}$.7, 31\arcdeg00\arcmin00\arcsec.0  & 4.9 ks\\
	   						    						     		    
0082540501  & 2002-Dec-09 & 20$^{\mathrm h}$50$^{\mathrm m}$18$^{\mathrm s}$.4, 30\arcdeg44\arcmin34\arcsec.3  & 12.6 ks\\
	   						    						     		    
0082540601  & 2002-Dec-11 & 20$^{\mathrm h}$49$^{\mathrm m}$02$^{\mathrm s}$.0, 30\arcdeg29\arcmin08\arcsec.6  & 11.5 ks\\
	   						    						     		    
0082540701  & 2002-Dec-13 & 20$^{\mathrm h}$47$^{\mathrm m}$45$^{\mathrm s}$.8, 30\arcdeg13\arcmin42\arcsec.9  & 13.7 ks\\

0405490101  & 2006-May-13 & 20$^{\mathrm h}$50$^{\mathrm m}$32$^{\mathrm s}$.2, 30\arcdeg11\arcmin00\arcsec.0  & 6.5 ks\\
	   						     						     		     
0405490201  & 2006-May-13 & 20$^{\mathrm h}$49$^{\mathrm m}$54$^{\mathrm s}$.2, 29\arcdeg42\arcmin25\arcsec.0  & 3.6 ks\\
\hline
\end{tabular}
\end{center}
\end{table}

\begin{table}
  \caption{Spectral parameters inferred from the combined XIS and EPIC data (see figure \ref{fig:XIS_Gau})}\label{tab:XIS_Gau}
\begin{center}
\begin{tabular}{lccc}
\hline \hline
  &  \multicolumn{2}{c}{XIS (1.3-5.0 keV)} & EPIC (1.5-5.0 keV) \\
\hline
 Parameter & \multicolumn{1}{c}{Bremss + 6Gauss} & \multicolumn{1}{c}{Bremss + 8Gauss} & \multicolumn{1}{c}{Bremss + 7Gauss} \\
\hline
N$\rm _H$ [10$^{20}$cm$^{-2}$]\dotfill &  \multicolumn{3}{c}{2.3 (fixed)} \\
$kT_e$ [keV]\dotfill &  0.61$^{+0.03}_{-0.02}$ & 0.59$^{+0.03}_{-0.04}$ & 0.51$\pm$0.01 \\
Line Center [eV]: & & \\
\ \ Mg He$\alpha$\dotfill  & 1341$\pm$1 & 1341$\pm$1 & $-$\\
\ \ Mg He$\beta$\dotfill  & 1572$^{+3}_{-4}$ & 1572$^{+5}_{-4}$ & 1555$^{+5}_{-9}$\\
\ \ Si He$\alpha$\dotfill  & 1848$\pm$1 &  1848$\pm$1 & 1854$\pm$1\\
\ \ Si He$\beta$\dotfill  & 2178$^{+6}_{-10}$ & 2176$^{+4}_{-10}$ & 2185$^{+5}_{-3}$\\
\ \ Si He$\gamma$\dotfill  & 2332$^{+24}_{-26}$ & 2324$\pm$20 & 2333$\pm$21\\
\ \ S He$\alpha$\dotfill  & 2440$^{+3}_{-4}$  &  2440$^{+2}_{-4}$ & 2445$^{+5}_{-2}$\\
\ \ S He$\beta$\dotfill  & $-$ &  2936$^{+49}_{-61}$ & 2920$^{+50}_{-55}$\\
\ \ Ar He$\alpha$ \dotfill & $-$ &  3128$\pm$39 & 3132$^{+26}_{-29}$\\
$\chi^2$/dof\dotfill & 1050/998 & 993/994 & 1214/932\\
\hline
\multicolumn{3}{l}{The errors are in the range $\Delta\chi^{2}<2.7$.}\\
\end{tabular}
\end{center}
\end{table}

\begin{table}
  \caption{Abundances for low-$kT_e$ component}\label{tab:shell}
  \begin{center}
    \begin{tabular}{lc}
       \hline 
      \hline
Element & Abundance (in units of solar) \\
\hline
 C\dotfill &  0.46 \\
 N\dotfill &  0.28 \\
 O\dotfill &  0.20 \\
 Ne\dotfill & 0.36 \\
 Mg\dotfill & 0.23 \\
 Si\dotfill &  0.29 \\
 S\dotfill &  0.18 \\
 Fe\dotfill &  0.25 \\
      \hline
    \end{tabular}
 \end{center}
\end{table}

\begin{table}
  \caption{Spectral parameters (see figure \ref{fig:FIBIEPIC})}\label{tab:FIBIEPIC}
  \begin{center}
    \begin{tabular}{lcc}
       \hline 
      \hline
Parameter & XIS & EPIC \\
      \hline
      N$\rm _H$ [10$^{20}$cm$^{-2}$]\dotfill & \multicolumn{2}{c}{2.3 (fixed)} \\
      High-$kT_e$ component: \\
      \ \ $kT_e$ [keV]\dotfill &  0.69$^{+0.04}_{-0.02}$  & 0.49$\pm$0.02 \\
      \ \ Abundance:\\
      \ \ \ \ O (=C=N)\dotfill &  \multicolumn{2}{c}{1.42 (fixed)}  \\
      \ \ \ \ Ne\dotfill       &  \multicolumn{2}{c}{1.25 (fixed)} \\
      \ \ \ \ Mg\dotfill       &  0.82$\pm$0.03 & 1.05 (fixed) \\
      \ \ \ \ Si\dotfill       &  3.14$^{+0.09}_{-0.10}$ & 4.57$^{+0.12}_{-0.07}$ \\
      \ \ \ \ S\dotfill        &  6.4$^{+0.5}_{-0.4}$ & 6.2$^{+0.4}_{-0.2}$\\
      \ \ \ \ Ar\dotfill       &  9.0$^{+4.0}_{-3.8}$ & 8.4$^{+2.5}_{-2.7}$\\
      \ \ \ \ Fe (=Ni)\dotfill &  \multicolumn{2}{c}{3.20 (fixed)}\\
      \ \ $\tau_{lower}$ [s cm$^{-3}$]\dotfill & \multicolumn{2}{c}{0 (fixed)} \\
      \ \ log($\tau_{upper}$ [s cm$^{-3}$])\dotfill & 10.73$\pm$0.04 & $<$12 \\
      Low-$kT_e$ component: \\
      \ \ $kT_e$ [keV]\dotfill & \multicolumn{2}{c}{0.28 (fixed)} \\
      \ \ Abundances\dotfill & \multicolumn{2}{c}{fixed\footnotemark[$*$]} \\
      \ \ $\tau_{lower}$ [s cm$^{-3}$]\dotfill & \multicolumn{2}{c}{0 (fixed)} \\
      \ \ log($\tau_{upper}$ [s cm$^{-3}$])\dotfill & \multicolumn{2}{c}{11.45 (fixed)}\\
      $\chi ^2$/dof\dotfill &  1096/1004  & 1020/912 \\
      \hline
\multicolumn{3}{l}{\footnotemark[$*$]{ Fixed to the values shown in table \ref{tab:shell}.}}\\
\multicolumn{3}{l}{Other elements are fixed to solar values.}\\
\multicolumn{3}{l}{The errors are in the range $\Delta\chi^{2}<2.7$.}\\
    \end{tabular}
 \end{center}
\end{table}

\begin{table}
  \caption{Spectral parameters inferred from the combined XIS data (see figure \ref{fig:Region_Gau})}\label{tab:Region_Gau}
\begin{center}
\begin{tabular}{lccc}
\hline \hline
 Parameter & Region-A & Region-B & Region-C \\
\hline
N$\rm _H$ [10$^{20}$cm$^{-2}$]\dotfill &  \multicolumn{3}{c}{2.3 (fixed)} \\
$kT_e$ [keV]\dotfill &  0.46$\pm$0.01 & 0.55$\pm$0.01 & 0.35$\pm$0.01 \\
Line Center [eV]: & & \\
\ \ Mg He$\alpha$\dotfill  & 1339$\pm$1 & 1341$^{+1}_{-2}$ & 1340$^{+3}_{-1}$\\
\ \ Mg He$\beta$\dotfill  & 1572$^{+18}_{-15}$ & 1580$^{+8}_{-3}$ & 1586$^{+6}_{-12}$\\
\ \ Si He$\alpha$\dotfill  & 1848$\pm$1 &  1849$\pm$1 & 1848$\pm$1\\
\ \ Si He$\beta$\dotfill  & 2172$^{+8}_{-4}$ & 2170$^{+10}_{-20}$ & 2160$^{+28}_{-33}$\\
\ \ Si He$\gamma$\dotfill  & 2304$^{+16}_{-17}$ & 2332$^{+48}_{-49}$ & 2348$^{+55}_{-54}$\\
\ \ S He$\alpha$\dotfill  & 2440$\pm$3  &  2438$^{+8}_{-4}$ & 2444$^{+24}_{-25}$\\
\ \ S He$\beta$\dotfill  & 2918$^{+66}_{-91}$ &  2909$^{+16}_{-90}$ & 2970$\pm$85\\
\ \ Ar He$\alpha$ \dotfill & 3104$^{+32}_{-28}$ &  3160$^{+66}_{-72}$ & $-$\\
$\chi^2$/dof\dotfill & 899/957 & 851/957 & 1214/932\\
\hline
\multicolumn{3}{l}{The errors are in the range $\Delta\chi^{2}<2.7$.}\\
\end{tabular}
\end{center}
\end{table}

\begin{table}
  \caption{Spectral parameters inferred from combined XIS data (see figure \ref{fig:Region_NEI})}\label{tab:Region_NEI}
  \begin{center}
    \begin{tabular}{lccc}
       \hline 
      \hline
 & Region-A &  Region-B &  Region-C \\
\hline
 Parameter & \multicolumn{2}{c}{two-component VPSHOCK} & one-component VPSHOCK\\
      \hline
      N$\rm _H$ [10$^{20}$cm$^{-2}$]\dotfill   & \multicolumn{3}{c}{2.3 (fixed)}\\
      High-$kT_e$ component: \\
      \ \ $kT_e$ [keV]\dotfill &  0.80$^{+0.01}_{-0.06}$ & 0.65$\pm$0.01 & 0.36$^{+0.02}_{-0.04}$ \\
      \ \ Abundance:\\
      \ \ \ \ C \dotfill & \multicolumn{2}{c}{linked to O abundance } & 0.79 (fixed)\\
      \ \ \ \ N \dotfill & \multicolumn{2}{c}{linked to O abundance} &   0.20 (fixed)\\
      \ \ \ \ O \dotfill &  1.38 (fixed) &  1.46 (fixed) & 0.19 (fixed)\\
      \ \ \ \ Ne\dotfill &  1.15 (fixed) &  1.36 (fixed) &  0.43 (fixed)\\
      \ \ \ \ Mg\dotfill &  1.52$^{+0.24}_{-0.11}$ &  1.69$\pm$0.02 &  0.37$^{+1.24}_{-0.02}$ \\
       \ \ \ \ Si\dotfill &  7.29$^{+0.17}_{-0.69}$ &  2.63$^{+0.04}_{-0.05}$ &  0.48$^{+2.29}_{-0.05}$ \\
      \ \ \ \ S\dotfill &   8.96$^{+0.33}_{-0.51}$ &  3.24$^{+0.24}_{-0.27}$ &  1.40$^{+0.81}_{-0.52}$\\
     \ \ \ \ Ar\dotfill &  8.5$^{+3.3}_{-3.9}$ &  4.1$^{+2.6}_{-2.8}$ &  $<$0.33\footnotemark[$\ddagger$] \\
      \ \ \ \ Fe (=Ni)\dotfill &  5.06 (fixed) & 2.44 (fixed) &  0.42 (fixed) \\
        \ \ $\tau_{lower}$ [s cm$^{-3}$]\dotfill & \multicolumn{3}{c}{0 (fixed)} \\
     \ \ log($\tau_{upper}$ [s cm$^{-3}$]) \dotfill &  11.39$^{+0.04}_{-0.03}$ & 11.68$\pm$0.03 &  11.11$^{+0.29}_{-0.15}$ \\
      Low-$kT_e$ component: \\
      \ \ $kT_e$ [keV]\dotfill &  \multicolumn{2}{c}{0.28 (fixed)} & $-$\\
      \ \ Abundances\dotfill & \multicolumn{2}{c}{fixed\footnotemark[$*$]} & $-$\\
       \ \ $\tau_{lower}$ [s cm$^{-3}$]\dotfill & \multicolumn{2}{c}{0 (fixed)} & $-$\\
       \ \ log($\tau_{upper}$ [s cm$^{-3}$])\dotfill &  \multicolumn{2}{c}{11.45 (fixed)} & $-$\\
$\chi ^2$/dof\dotfill & 550/980  & 565/973 &  583/952\\
      \hline
\multicolumn{4}{l}{\footnotemark[$*$]{ Fixed to the values shown in table \ref{tab:shell}.}}\\
\multicolumn{4}{l}{\footnotemark[$\dagger$]{Estimated by fixing all other parameters. }}\\
\multicolumn{4}{l}{Other elements are fixed to solar values.}\\
\multicolumn{4}{l}{The errors are in the range $\Delta\chi^{2}<2.7$.}\\
   \end{tabular}
 \end{center}
\end{table}

\twocolumn

\end{document}